\documentclass[runningheads]{llncs}

\usepackage{amsmath}
\usepackage{tikz}
\usepackage{dblfloatfix}
\usepackage{ifthen}
\usepackage{multirow}
\usepackage{algorithm}
\usepackage{algorithmic}
\usepackage{graphicx}
\usepackage{amssymb}
\usepackage{xspace}
\usepackage{booktabs}
\usepackage{subcaption}
\usepackage{versions}
\usepackage[table]{colortbl}
\usetikzlibrary{shapes,arrows}
\usetikzlibrary{calc}
\usetikzlibrary{backgrounds, positioning, fit}
\usetikzlibrary{shapes,arrows}
\usetikzlibrary{backgrounds, positioning, fit}
\usepackage{pgfplots}
\usepackage[inline]{enumitem}
\usepackage{booktabs} 
\usepackage{siunitx}

\usetikzlibrary{decorations.shapes}
\tikzset{decorate sep/.style 2 args=
{decorate,decoration={shape backgrounds,shape=circle,shape size=#1,shape sep=#2}}}
\usetikzlibrary{decorations.pathreplacing}

\makeatletter
\tikzset{
    database/.style={
        path picture={
            \draw (0, 1.5*\database@segmentheight) circle [x radius=\database@radius,y radius=\database@aspectratio*\database@radius];
            \draw (-\database@radius, 0.5*\database@segmentheight) arc [start angle=180,end angle=360,x radius=\database@radius, y radius=\database@aspectratio*\database@radius];
            \draw (-\database@radius,-0.5*\database@segmentheight) arc [start angle=180,end angle=360,x radius=\database@radius, y radius=\database@aspectratio*\database@radius];
            \draw (-\database@radius,1.5*\database@segmentheight) -- ++(0,-3*\database@segmentheight) arc [start angle=180,end angle=360,x radius=\database@radius, y radius=\database@aspectratio*\database@radius] -- ++(0,3*\database@segmentheight);
        },
        minimum width=2*\database@radius + \pgflinewidth,
        minimum height=3*\database@segmentheight + 2*\database@aspectratio*\database@radius + \pgflinewidth,
    },
    database segment height/.store in=\database@segmentheight,
    database radius/.store in=\database@radius,
    database aspect ratio/.store in=\database@aspectratio,
    database segment height=0.1cm,
    database radius=0.25cm,
    database aspect ratio=0.35,
}
\makeatother
\newcommand{\makeparafit}{\looseness=-1}

\newcommand{\verifier}{$\mathcal{V}$\xspace}
\newcommand{\prover}{$\mathcal{P}$\xspace}
\newcommand{\malprover}{$\mathcal{P}_{mal}$\xspace}
\newcommand{\serverone}{$\mathcal{S}_{1}$\xspace}
\newcommand{\servertwo}{$\mathcal{S}_{2}$\xspace}
\newcommand{\adv}{$\mathcal{A}dv$\xspace}

\newcommand{\attclf}{$f_{att}$\xspace}
\newcommand{\data}{\mathbb{D}\xspace}

\newcommand{\census}{\textsc{CENSUS}\xspace}
\newcommand{\racecensus}{\textsc{CENSUS-R}\xspace}
\newcommand{\sexcensus}{\textsc{CENSUS-S}\xspace}
\newcommand{\boneage}{\textsc{BONEAGE}\xspace}
\newcommand{\arxiv}{\textsc{ARXIV}\xspace}
\newcommand{\model}{$\mathcal{M}^{\theta}$\xspace}
\newcommand{\modelcheck}{$\mathcal{M}_{p}$\xspace}
\newcommand{\firstlayer}{$\mathcal{M}_{p}^1$\xspace}
\newcommand{\firstlayerdelta}{$\mathcal{M}_{p}^1+\delta$\xspace}
\newcommand{\modelmpc}{$\mathcal{M}_{2pc}$\xspace}

\newcommand{\proverdataorig}{$\mathcal{D}_{\mathcal{P}}$\xspace}
\newcommand{\proverdata}{$\mathcal{D}^{tr}_{\mathcal{P}}$\xspace}
\newcommand{\verificationdata}{$\mathcal{D}^{ver}_{\mathcal{P}}$\xspace}
\newcommand{\verifierdata}{$\mathcal{D}_{\mathcal{V}}$\xspace}
\newcommand{\verifiertraindata}{$\mathcal{D}^{tr}_{\mathcal{V}}$\xspace}
\newcommand{\verifiertestdata}{$\mathcal{D}^{test}_{\mathcal{V}}$\xspace}
\newcommand{\preq}{$p_{req}$\xspace}
\newcommand{\pset}{$\overline{\textbf{p}}$\xspace}
\newcommand{\Nspotcheck}{$\mathcal{N}_{spchk}$\xspace}
\newcommand{\Naccepts}{\mathcal{N}_{a}\xspace}
\newcommand{\Pfallback}{\mathbb{P}_{crpt}\xspace}
\newcommand{\Pinf}{\mathbb{P}_{inf}\xspace}
\newcommand{\Pspotcheck}{\mathbb{P}_{spchk}\xspace}
\newcommand{\Nrejects}{\mathcal{N}_{rej}\xspace}
\newcommand{\Nmodels}{\mathcal{N}_{m}\xspace}

\newcommand{\Ntotal}{\mathcal{N}\xspace}\newcommand{\CostInf}{\omega_{inf}\xspace}
\newcommand{\CostCrypt}{\omega_{crpt}\xspace}
\newcommand{\CostCryptComp}{\omega_{crpt}^{comp}\xspace}
\newcommand{\CostCryptComm}{\omega_{crpt}^{comm}\xspace}

\newcommand{\dc}{\texttt{DistCheck}\xspace}

\newcommand{\modelmpcshare}[1]{[\mathcal{M}_{2pc}]^{#1}}

\newcommand{\proverdatashare}[1]{[\mathcal{D}^{tr}_{\mathcal{P}}]^{#1}}

\newcommand{\maliciousdatashare}[1]{[\mathcal{D}']^{#1}}

\includeversion{topics}

\excludeversion{full}

\newcommand{\Hypar}[1]{\noindent\textbf{#1.}}
\newcommand{\Itpar}[1]{\noindent\textit{#1.}}

\includeversion{preprint}
\excludeversion{submit}

\begin{document}
\title{Attesting Distributional Properties of Training Data for Machine Learning}

\titlerunning{Attesting Distributional Properties of Training Data for Machine Learning}

\author{Vasisht Duddu\inst{1} \and
Anudeep Das\inst{1} \and
Nora Khayata\inst{2} \and
Hossein Yalame\inst{2} \and
Thomas Schneider\inst{2} \and 
N. Asokan\inst{1}}

\authorrunning{Duddu et al.}

\institute{University of Waterloo \and
Technical University of Darmstadt\\
\email{\{vasisht.duddu, a38das\}@uwaterloo.ca, \{khayata, yalame, schneider\}@encrypto.cs.tu-darmstadt.de, asokan@acm.org}}

\maketitle

\begin{abstract}
The success of machine learning (ML) has been accompanied by increased concerns about its trustworthiness. Several jurisdictions are preparing ML regulatory frameworks. One such concern is ensuring that model training data has desirable \emph{distributional properties} for certain sensitive attributes. For example, draft regulations indicate that model trainers are required to show that training datasets have specific distributional properties, such as reflecting the diversity of the population. 
We propose the novel notion of \textit{ML property attestation} allowing a prover (e.g., model trainer) to demonstrate relevant properties of an ML model to a verifier (e.g., a customer) while preserving the confidentiality of sensitive data. We focus on the attestation of distributional properties of training data \emph{without revealing the data}. We present an effective hybrid property attestation combining property inference with cryptographic mechanisms. 
\begin{preprint}
\footnote{An abridged version of this paper appears in the 29$^{th}$ European Symposium on Research in Computer Security (ESORICS), 2024}
\end{preprint}

\end{abstract}

\section{Introduction}\label{sec:introduction}

Machine learning (ML) models are being deployed for a wide variety of critical real-world applications such as criminal justice, healthcare, and finance. 
This has raised several trustworthiness concerns~\cite{SoKMLPrivSec}. There are indications that future regulations will require ML model trainers to account for these concerns~\cite{congress,eur}.  One such concern is to 
ensure that the training data has desirable \emph{distributional properties} with respect to characteristics such as gender or skin color, e.g., the proportion of training data records with a certain attribute value such as skin-tone=black is consistent with the proportion in the population at large. 
Forthcoming regulation may require model owners to demonstrate such \emph{distributional equity} in their training data, showing that distributional properties of certain training data attributes fall within ranges specified by regulatory requirements: e.g., the draft \emph{Algorithmic Accountability Act} bill~\cite{congress} requires operators of automated decision systems to keep track of ``the representativeness of the dataset and how this factor was measured including \ldots the distribution of the population'' (cf.~\cite[\S7.C.(i)]{congress}). The European Parliament's proposed AI act~\cite{eur} stipulates that ``datasets \ldots 
shall have the appropriate statistical properties, including, where applicable, as regards the persons or groups of persons on which the high-risk AI system is intended to be used'' (cf.~\cite[Art. 10.3]{eur}). This ensures that there are no errors arising from population misalignment, i.e., the model does not accurately represent the target population due to distribution shifts between training data and data seen in the real-world~\cite{10136159}. 

These regulations do not (yet) spell out technical mechanisms for verifying compliance. In this paper, we introduce the notion of \textit{ML property attestation}, which are technical mechanisms by which a \emph{prover} (e.g., a model trainer) can demonstrate relevant properties about the model to a \emph{verifier} (e.g., regulatory agency or a customer purchasing the trained model). Properties of interest may correspond to either training (relating to the model, its training data, or the training process) or inference (e.g., relating to the inference process, or binding the model to its inputs and/or outputs).
We focus on \emph{distributional property attestation}, proving distributional properties of a training dataset to the verifier.

A na\"ive approach for distributional property attestation is to have the prover reveal the training data to the verifier. But this na\"ive approach may not be legally or commercially viable, given the sensitivity and/or business value of the training data. We identify four requirements for property attestation: be 
\begin{enumerate*}[label=\roman*),itemjoin={,\xspace}]
\item \textit{effective}
\item \textit{efficient}
\item \textit{confidentiality-preserving}
\item \textit{adversarially robust}.
\end{enumerate*}
Simultaneously meeting all of them is challenging. The natural approaches of using trusted execution environments (TEEs), or cryptographic protocols, like secure two-party computation (2PC) and zero knowledge proofs (ZKPs), either impose deployability hurdles or incur excessive overheads.

An interesting alternative is to adapt \emph{property inference attacks} which infer distributional properties of training datasets~\cite{ateniesepropinf}. Here, the verifier runs a property inference protocol against the prover's model. Some proposed property inference attacks make strong, unrealistic, assumptions about adversary capabilities, e.g., whitebox model access~\cite{suripropinf}. We argue that such assumptions are reasonable in our attestation setting where provers and verifiers are \emph{incentivized to collaborate} to complete the attestation. Given the changed adversary model, property inference techniques need to be adapted to ensure adversarial robustness against malicious provers.
\setlist{nosep,topsep=-\parskip}
\textbf{Our main contributions are as follows:}
\kern-\parskip\begin{enumerate}[leftmargin=*]
\item the novel notion of \textit{ML property attestation}, and desiderata for effective mechanisms to attest distributional properties of training data (\S\ref{sec:problem}), and 
\item a \emph{hybrid attestation mechanism}\footnote{Code: \url{https://github.com/ssg-research/distribution-attestation}.}, combining a property inference attack technique with 2PC (\S\ref{sec:approach}), and extensive empirical evaluation showing its effectiveness (\S\ref{sec:setup} and \S\ref{sec:evaluation}).
\end{enumerate}\kern-\parskip

\section{Background}\label{sec:background}

We first summarize ML notations, distributional properties of training data, property inference attacks, and secure multi-party computation (MPC).

\Hypar{ML Notations}\label{sec:backml} Consider a data distribution~$\data$ and a training dataset $\mathcal{D}_{tr}\sim\data$ with $\mathcal{D}_{tr} =\{x_i,y_i\}_{i}^N$ where the $i^{th}$ tuple consists of a vector of \emph{attributes} $x_i$ and its classification label~$y_i$. 
An ML classification model is a function $\mathcal{M}^{\theta} :x \rightarrow y$, parameterized by the model parameters $\theta$, which maps input features $x$ to their corresponding classification label $y$. During training, $\theta$ is iteratively updated by penalizing the model for incorrectly predicting $y$ given $x \in \mathcal{D}_{tr}$. During inference, an input $x'$ to \model gives the prediction $\mathcal{M}^{\theta}(x')$. We omit $\theta$ in \model.

\Hypar{Distributional Properties of Training Data}\label{sec:propinfback} 
We borrow the definition for the distributional property of training data $\mathcal{D}_{tr}$ from Suri and Evans~\cite{suripropinf}. 
A distributional property is the ratio of an indicator function, counting different data records applied to a dataset (uniformly sampled from a distribution) with a specific attribute value (e.g., males), and total number of data records or number of records with other attribute value (e.g., females). Examples include the ratio of males to females or whites to non-whites in tabular or image datasets, or the average node degree and clustering coefficient for graph data~\cite{suripropinf,suri2022dissecting}.

\Hypar{Property Inference Attacks}\label{sec:propinfback} Property inference attacks allow an adversary \adv to infer such distributional properties about \emph{sensitive} attributes in the data distribution $\data$ (e.g., ratio of males/females) using access to the model $\mathcal{M}$~\cite{ganjupropinf,suri2022dissecting,suripropinf,zhangpropinf,ganpropinf,PasquiniSplitInf,infAttGraphs}. The attack assumes that a model trainer and \adv have access to $\data$ and sampling functions $\mathcal{G}_0$ and $\mathcal{G}_1$ which transform $\data$ to obtain a sub-distribution satisfying a particular property. For instance, $\mathcal{G}_0(\data)$ indicates 80\% males and 20\% females while $\mathcal{G}_1(\data)$ indicates 50\% males and 50\% females. Given models $\mathcal{M}_0$ and $\mathcal{M}_1$ trained on datasets sampled from these sub-distributions $\mathcal{G}_0(\data)$ and $\mathcal{G}_1(\data)$, \adv infers whether $\mathcal{M}$ was trained on $\mathcal{G}_0(\data)$ or $\mathcal{G}_1(\data)$ (i.e., $\mathcal{D}_{tr}$ has 80\% males or 50\% males).

\Hypar{MPC}\label{sec:cryptoback} This cryptographic protocol allows mutually distrusting parties to jointly compute a function on their private inputs, such that nothing beyond the output is leaked~\cite{lindell2020secure}.
MPC has been adopted to a wide range of applications, including financial services~\cite{CTRSA:AtapoorSA22} and privacy-preserving machine learning~\cite{knott2021crypten}.
We make use of secure two-party computation (2PC), a form of MPC with one dishonest party. Dishonest parties can be either semi-honest (follow the protocol but try to infer the other party's inputs) or malicious (deviate from the protocol, e.g., to break correctness). While maliciously secure MPC protocols are more secure, they come with higher computation and communication costs~\cite{zheng2021cerebroMalicious}. For real-world applications, semi-honest security guarantees are often sufficient and give baseline performance numbers~\cite{mpcallianceAlliance,berkeleyDeployments}.

\section{Problem Statement}\label{sec:problem}

We first present the notion of \emph{ML property attestation} followed by the system and adversary models to attest \emph{distributional properties}. We then identify desiderata for distributional property attestation mechanisms.

\Hypar{Property Attestation} 
These are technical mechanisms using which a prover \prover (e.g., a model trainer) can prove to a verifier \verifier (e.g., potential customer purchasing the model or regulator) that a certain property about the model holds. For example, distributional property attestation can prove that the proportion of records having a specific value of a given attribute in \prover's training dataset~\proverdataorig~$\sim$~$\data$ meets the value \preq expected by \verifier. Both \prover and \verifier know \preq. Hereafter, we focus on distributional property attestation.

\Hypar{System and Adversary Models} We assume that the distributional property for the attribute of interest can take a set of $n$ possible values \pset=\{$p_0$,\ldots,$p_n$\} (e.g., proportion of females in the dataset). Following the literature on property inference attacks, we assume that both \prover and \verifier know $\data$~\cite{suripropinf,zhangpropinf,fedLearningAttInf,suri2022dissecting}. 
\proverdataorig is split into \proverdata and \verificationdata: \proverdata is used to train \modelcheck with some property, \verificationdata, which is not known by \verifier, is used for evaluating attestation to simulate what \verifier is likely to see in practice. \verifier has their own dataset \verifierdata$\sim\data$. \verifierdata is split into a training dataset (\verifiertraindata) used for building attestation mechanism and test dataset (\verifiertestdata) to locally evaluate the mechanism.

The goal of \prover, who has trained a model \modelcheck on \proverdata, is to succeed in property attestation to comply with regulation. \verifier's goal is to ensure that attestation succeeds if \proverdata meets \preq even if \prover tries to fool the attestation process.
We assume that \prover has given \verifier whitebox access to \modelcheck. This is reasonable since \prover is incentivized to co-operate with \verifier to complete the attestation successfully. However, \prover does not want to disclose \proverdata to~\verifier for confidentiality/privacy.

\Hypar{Requirements} A property attestation mechanism must be:
\begin{enumerate}[label=\textbf{R\arabic*},leftmargin=*]
\item \label{req1}\textbf{Confidentiality-preserving:} \verifier \begin{preprint}(or any other entity)\end{preprint} learns no~additional~information about~\proverdata;
\item \label{req2}\textbf{Effective:} correctly identify if \proverdata meets \preq, with acceptably low false accepts (FA)/rejects (FR)
\item \label{req3}\textbf{Adversarially robust:} meet \ref{req2}, with respect to FA, even if \prover misbehaves
\item \label{req4}\textbf{Efficient:} impose an acceptable computation and communication overhead.
\end{enumerate}

\section{Distributional Property Attestation Mechanisms}\label{sec:approach}

Property attestation by simply revealing \proverdata to \verifier violates \ref{req1} and is susceptible to manipulations by a malicious \prover (\malprover). We discuss three different property attestation mechanisms satisfying \ref{req1} by design and examine~\ref{req2}-\ref{req4} for each mechanism: inference-based attestation, cryptographic attestation using MPC, and a hybrid attestation combining the benefits of both.

\begin{figure}[h]
\begin{center}
\resizebox{\textwidth}{!}{
\begin{tikzpicture}
\node [rectangle,draw,fill=black,thick,minimum width=1.5cm, minimum height=0.55cm,xshift=2cm,yshift=5cm] at (0,0) (verifier1) {\textcolor{white}{Verifier (\verifier)}};
\node [above of = verifier1, yshift=-0.4cm] (prep) {\underline{Preparation (Train \attclf)}};

\node [below of = verifier1, xshift=-2cm, rectangle,draw,thick,minimum width=1.5cm, minimum height=0.75cm, align=center] (m1preq) {$\mathcal{M}^1_{p_{req}}$};
\node [right of = m1preq, xshift=1.4cm, rectangle,draw,thick,minimum width=1.5cm, minimum height=0.75cm, align=center] (mnpreq) {$\mathcal{M}^n_{p_{req}}$};
\draw [dotted, ultra thick](1,4) -- (1.5,4) ;

\begin{scope}[on background layer]
    \node (con1) [fit=(m1preq) (mnpreq), draw=gray,dashed, fill= yellow!8, ultra thick, rounded corners, inner sep=0.1cm] {};
\end{scope}

\node [left of = con1,xshift=-1.9cm, align=center] (result) {\tiny Shadow Models\\\tiny\preq};

\node [below of = m1preq, yshift=-0.2cm, rectangle,draw,thick,minimum width=1.5cm, minimum height=0.75cm, align=center] (m1p1) {$\mathcal{M}^1_{p_{1}}$};
\node [right of = m1p1, xshift=1.4cm, rectangle,draw,thick,minimum width=1.5cm, minimum height=0.75cm, align=center] (mnp1) {$\mathcal{M}^n_{p_{1}}$};
\draw [dotted, ultra thick](1,2.75) -- (1.5,2.75) ;
\draw [dotted, ultra thick](0,2.1) -- (0,2.3) ;

\node [below of = m1p1, yshift=-0.2cm, rectangle,draw,thick,minimum width=1.5cm, minimum height=0.75cm, align=center] (m1pn) {$\mathcal{M}^1_{p_{n}}$};
\node [right of = m1pn, xshift=1.4cm, rectangle,draw,thick,minimum width=1.5cm, minimum height=0.75cm, align=center] (mnpn) {$\mathcal{M}^n_{p_{n}}$};
\draw [dotted, ultra thick](1,1.65) -- (1.5,1.65) ;
\draw [dotted, ultra thick](2.5,2.1) -- (2.5,2.3) ;

\begin{scope}[on background layer]
    \node (con2) [fit=(m1p1) (mnp1) (m1pn) (mnpn), draw=gray,dashed, fill= yellow!8, ultra thick, rounded corners, inner sep=0.1cm] {};
\end{scope}

\node [left of = con2,xshift=-1.9cm, align=center] (result) {\tiny Shadow Models\\\tiny!\preq};

\node [right of = con2, xshift=2.5cm, yshift=0.75cm, rectangle,draw,thick,minimum width=1.5cm, minimum height=0.75cm, align=center] (attclf0) {\attclf};


\node [rectangle,draw,thick,minimum width=1.5cm,xshift=6cm, fill=black, minimum height=0.55cm, right of=verifier1] (prover) {\textcolor{white}{Prover (\prover)}};
\node [right of=prover,xshift=3cm,rectangle,draw,fill=black,thick,minimum width=1.5cm, minimum height=0.55cm] (verifier) {\textcolor{white}{Verifier (\verifier)}};
\node [above of = prover, yshift=-0.4cm,xshift=2cm] (attext) {\underline{Attestation}};
\draw [dashed, ultra thick](7,6) -- (7,1) {};

\node[database,below of = prover, yshift=-0.2cm,label=above:$\mathcal{D}_{\mathcal{P}}$] (data) {};

\node [below of = data, rectangle,draw,thick,minimum width=1.5cm, minimum height=0.75cm] (localmodel) {\modelcheck};
\node [right of = localmodel, xshift=3cm, rectangle,draw,thick,minimum width=1.5cm, minimum height=0.75cm, align=center] (attclf2) {\attclf};

\draw[->,ultra thick] (data.south) -- node[anchor=west, align=center] {\em\footnotesize Train w/ \proverdata} (localmodel.north);
\draw[->,ultra thick] (localmodel.east) -- node[anchor=east, align=center] {\em\footnotesize } (attclf2.west);
\node [below of = attclf2,yshift=-0.3cm] (result) {\preq or !\preq?};
\draw[->,ultra thick] (attclf2.south)--(result.north);

\draw[->,ultra thick] (con1.east)-| node[anchor=south, align=center] {\em\footnotesize label=1}(attclf0.north);
\draw[->,ultra thick] (con2.east)-| node[anchor=north, align=center] {\em\footnotesize label=0}(attclf0.south);
\end{tikzpicture}
}
\end{center}
\caption{\textbf{Inference-based Attestation}: During preparation, \verifier trains \attclf using the first layer parameters of models trained on the training data \proverdata with \preq ($\{\mathcal{M}^i_{p_{req}}\}_{i=1}^{\Nmodels}$) and !\preq ($\{\mathcal{M}^i_{!p_{req}}\}_{i=1}^{\Nmodels}$). During attestation, \verifier uses first layer parameters of \modelcheck to attest if it was indeed trained on \proverdata with \preq or not.} 
\label{fig:infatt}
\vspace{-0.6cm}
\end{figure}
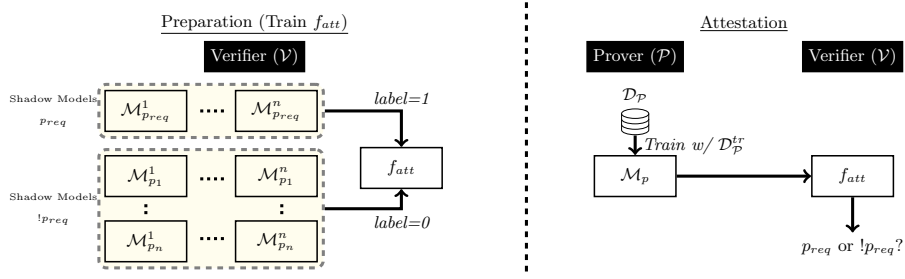

\Hypar{Inference-based Attestation} \label{sec:infatt} Recall that property inference attacks infer statistical properties of training data given access to the victim's model. Hence, these attacks can be adapted for property attestation. Unlike the attack where whitebox model access to \adv is a strong assumption, \prover and \verifier have an incentive to collaborate to complete the attestation successfully, making whitebox access reasonable. 
However, directly applying property inference attacks is not possible as there are differences between the two settings (inference attack vs. attestation) in terms of their:
\begin{itemize}[leftmargin=*]
    \item \textbf{objective}: the attack distinguishes between two property values while attestation requires differentiating \preq from all others (!\preq). 
    \item \textbf{requirement}: attestation has the additional requirement of robustness \ref{req3}, i.e., resist \malprover's attempts to fool \verifier.
\end{itemize}

We show how property inference attacks can be adapted to attestation and describe the inference-based attestation below.

\Itpar{Method} Given access to \modelcheck, \verifier uses an attestation classifier (\attclf) to attest if \proverdata satisfies \preq using the first layer parameters of \modelcheck as input to \attclf. The first layer parameters are more effective to capture distributional properties for successful property inference than subsequent layers~\cite{suripropinf}.
To train \attclf, \verifier uses \verifiertraindata and generates multiple sub-distributions \{$\mathcal{G}_0(\data), \ldots{}, \mathcal{G}_n(\data)$\} corresponding to property values in \pset and samples datasets \{$\mathcal{D}_0, \ldots,\mathcal{D}_n$\}. In practice, this is done by sampling datasets multiple times with different properties from \verifierdata. For each dataset and property value, \verifier trains $\Nmodels$ ``shadow models'' \{\{$\mathcal{M}^i_0\}_{i=1}^{\Nmodels}$, \ldots, \{$\mathcal{M}^i_n\}_{i=1}^{\Nmodels}$\}. These mimic the 
models that \verifier could encounter during attestation.

\verifier trains \attclf using the first layer parameters of models trained on \proverdata with \preq ($\{\mathcal{M}^i_{p_{req}}\}_{i=1}^{\Nmodels}$) and !\preq ($\{\mathcal{M}^i_{!p_{req}}\}_{i=1}^{\Nmodels}$). \verifier uses \verifiertestdata for evaluating \attclf. Attestation effectiveness is evaluated using \verificationdata. 
We present a visualization of inference-based attestation in Figure~\ref{fig:infatt}.

\Hypar{Cryptographic Attestation}\label{sec:crypto} Property attestation can be securely achieved using cryptographic protocols (e.g., MPC, ZKPs) by proving that (a) \proverdata meets \preq (\dc), and (b) \modelcheck was trained on \proverdata to ensure that a misbehaving \prover does not change \proverdata after (a) (Figure~\ref{fig:cryptatt}). We use 2PC due to their practicality (see \S\ref{sec:discussions} for discussion on alternative approaches).

\Itpar{Assumptions} \prover may deceive \verifier about \preq, acting maliciously. However, \verifier, interested in purchasing \modelcheck, has no incentive to cheat, but may seek additional details about \proverdata. Thus, we assume \verifier behaves semi-honestly.

\Itpar{Setup} To account for \malprover, we could use malicious two-party protocols directly between \prover and \verifier, which is prohibitively expensive. 
Instead, \prover and \verifier rely on secure outsourced computation to independent and non-colluding servers \serverone and \servertwo as done in prior work~\cite{Riazi2018ChameleonAH} and in practical deployments~\cite{mozillaDeployment}. \serverone and \servertwo can be instantiated by different companies, which according to data protection laws must protect user data, thus cannot share their data with each other. Outsourcing also allows to flexibly instantiate the cryptographic protocol, i.e., our construction generalizes to 2PC/MPC or ZKP protocols.

\begin{figure}[h]
\begin{center}
\resizebox{.65\textwidth}{!}{
\begin{tikzpicture}

\node [rectangle,draw,thick,minimum width=1.5cm, fill=black, minimum height=0.55cm] at (0,0) (prover) {\textcolor{white}{Prover (\prover)}};

\node [right of = prover, xshift=3cm, yshift=0.65cm, rectangle,draw,thick,minimum width=1.5cm, minimum height=0.55cm] (s1) {\serverone};
\node [below of = s1, yshift=-0.3cm, rectangle,draw,thick,minimum width=1.5cm, minimum height=0.55cm] (s2) {\servertwo};

\node [right of=s2,yshift=0.65cm,xshift=3cm,rectangle,draw,fill=black,thick,minimum width=1.5cm, minimum height=0.55cm] (verifier) {\textcolor{white}{Verifier (\verifier)}};

\begin{scope}[on background layer]
    \node (con1) [fit=(s1) (s2), draw=gray,dashed, fill= yellow!8, ultra thick, rounded corners, inner sep=0.2cm, label={[align=center]Independent Non-Colluding Servers\\for outsourced semi-honest 2PC}] {};
\end{scope}

\draw[<->,ultra thick] (s1.south)-- node[anchor=west, align=center] {\em\footnotesize 2PC}(s2.north);
\draw[->,ultra thick] (prover.east) -- node[anchor=north, align=center] {\tiny \proverdata Secret\\ \tiny Shares}(con1.west);
\draw[->,ultra thick] (con1.east) -- node[anchor=north, align=center] {\tiny Output Secret\\\tiny Shares}(verifier.west);

\end{tikzpicture}
}
\end{center}
\caption{\textbf{Cryptographic Attestation}: \prover sends the secret shares of the training data \proverdata to \serverone and \servertwo. The servers securely compute ``DistCheck'' for \proverdata and train \modelmpc on \proverdata with their secret shares using 2PC. The output shares are then sent to \verifier for reconstructs the outputs.} 
\label{fig:cryptatt}
\vspace{-0.6cm}
\end{figure}

\Itpar{Method} We consider secret sharing over a ring with $Q$ elements where a secret input $x$ is split into two shares, $x_1$ and $x_2$ such that $x=x_1+x_2 \mod Q$~\cite{knott2021crypten}. Each share $x_1$ and $x_2$ looks random, i.e., given only one share, one cannot learn any information about the secret. For (a), \dc computes the distributional property directly over \proverdata and comparing with \preq. Here, \prover sends secret shares of \proverdata to \serverone and \servertwo who jointly perform \dc by running secure accumulation and comparison using 2PC. For (b), \prover sends secret shares of the initial model weights to obtain \modelcheck to \serverone and \servertwo. Together with the previously obtained shares of \proverdata, \serverone and \servertwo jointly run secure training. \serverone and \servertwo send the resulting secret shares of \dc and the final model parameters of the trained model \modelmpc to \verifier who adds the received shares to get the results of \dc and the trained model weights. The \textit{correctness} property of 2PC convinces \verifier that both \dc and training are run correctly on \proverdata. We implement secure computation between \serverone and \servertwo using CrypTen~\cite{knott2021crypten}, a framework for efficient privacy-preserving ML that supports one semi-honest corruption for two parties (see Appendix~\ref{app:crypto} for more details, security and correctness of the protocols).

\Hypar{Hybrid Property Attestation}\label{sec:hybrid} Cryptographic attestation is costly, while inference-based attestation can have unacceptably high false acceptance or false rejected rates (FAR or FRR respectively) (Table~\ref{tab:effectiveness}). Relying solely on either is inadequate. Therefore, we propose a hybrid attestation scheme that first uses inference-based attestation with cryptographic attestation as a fallback. Depending on the application, \verifier can fix an acceptably low FAR or FRR:
\begin{itemize}[leftmargin=*]
\item \textbf{Fixed FAR.} For accepted provers (\prover{s}), no further action is needed. If the inference-based attestation fails (FR), \prover{s} can request re-evaluation with cryptographic attestation.
\item \textbf{Fixed FRR.} If inference-based attestation is rejected, there is no provision for re-appeal since FRR is low. For accepted \prover{s}, \verifier may do a random ``spot-check'' using cryptographic attestation.
\end{itemize}

\Itpar{Assumptions} We assume FAR and FRR are fixed at $5\%$. 
Additionally, \verifier uses \modelcheck for inference-based attestation and \modelmpc is obtained by 2PC training. We assume that \prover shares the hyperparameters for training \modelcheck to obtain \modelmpc to be perfectly equivalent. This can be done by a fidelity check, i.e., sending arbitrary inputs and ensuring outputs from \modelcheck and \modelmpc are equal. 

\Itpar{Method} While hybrid attestation is straight-forward for fixed FAR, we describe the methodology for random spot-checks of accepted \prover{s} for fixed FRR. Let $z$ be the total FA on \verifiertestdata from the inference-based attestation and $\Naccepts$ denote the number of accepted \prover{s}. 
Knowing $z$, \verifier can randomly sample \Nspotcheck spot-checks where $z \leq$ \Nspotcheck $\leq \Naccepts$.
\verifier then uses cryptographic attestation to eliminate any FA in the sampled set thus reducing the overall FAR. To compute the new FAR, we first compute the probability of finding $t$ FAs from the sample of \Nspotcheck \prover{s}. We model the probability distribution over FA as hypergeometric distribution which computes the likelihood of selecting $t$ FAs in a sample of \Nspotcheck from a population of $z$ falsely accepted \prover{s} without replacement: $\mathbb{P}(T = t) = {\binom{z}{t}\binom{\Naccepts-z}{\mathcal{N}_{spchk}-t}}/{\binom{\Naccepts}{\mathcal{N}_{spchk}}}$, where $t \in [0, z]$.  We compute the effective \#FAs as $\#FA_{new} = \#FA_{old} - t'$ where $t' = argmax_{t\in[0,z]} \mathbb{P}(T = t)$. \Nspotcheck will determine the FAR and cost incurred.

\section{Experimental Setup}\label{sec:setup}

We describe the different datasets used and corresponding model architectures, followed by the metrics used for evaluation.

\Hypar{Datasets and Model Architecture}\label{sec:datasets} We use the datasets, properties, and model architectures same as in prior work on property inference attacks~\cite{suripropinf}. 


\noindent\textbf{\underline{\boneage}} is an image dataset which contains X-Ray images of hands, with the task being predicting the patient’s age in months. The dataset is converted to a binary classification task for classifying the age of the patient. We focus on the ratios of the females the property of interest. We consider the following permissible ratios (\pset): [``0.2'' - ``0.8'']. Here, the sensitive attribute is implicit as part of the metadata.

\noindent\textbf{\boneage model} is a pre-trained DenseNet~\cite{densenet} model for feature extraction of the images, followed by a three-layer network of size [128, 64, 1] for classification with ReLU activations. We train the model for 100 epochs with a batch size of 8192, learning rate of 0.001, and weight decay of 1e-4.

\noindent\textbf{\underline{\arxiv}} is a directed graph dataset representing citations between computer science ArXiv papers. The classification task is to predict the subject area for the papers. The property considered for attestation is the mean node-degree of the graph dataset. We use the following permissible ratios (\pset): [``9'' - ``17'']. The graph dataset is sampled to satisfy a specific mean-node degree which is implicitly included in the dataset. 

\noindent\textbf{\arxiv model} is a four-layer graph convolutional network which maps the input graph data to low dimensional embedding for node classification tasks. The graph convolution layer sizes are [256, 256, 256, and 40] with ReLU activation. We use dropout with 0.5 drop probability. We train the model for 100 epochs with a learning rate of 0.01, and weight decay of 5e-4.

\noindent\textbf{\underline{\census}} is a tabular dataset which consists of several categorical and numerical attributes like age, race, education level. The classification task is to predict whether an individual’s annual income exceeds 50K. However, we have two variants of this dataset based on the property: (a) \racecensus which considers the distribution of whites and (b) \sexcensus which considers the distribution of females in the dataset. For both, we consider the following permissible ratios (\pset): [``0.0'' - ``1.0'']. Both \sexcensus and \racecensus explicitly include the sensitive attributes in the dataset. 

\noindent\textbf{\census model} is a three layer deep neural network with the hidden layer dimensions: [32, 16, 8] with ReLU activation function. 
We train the model for 100 epochs with a learning rate of 0.01, batch size of 64, and no weight decay.

Our \attclf is based on permutation invariant networks, based on  DeepSets architecture~\cite{deepsets}, used in the property inference literature~\cite{suripropinf,ganjupropinf}. The classifier generates a representation of the input model parameters independent of the ordering of the neurons in a layer. Suri and Evans~\cite{suripropinf} suggest that the first layer captures distributional properties better than subsequent layers. Hence, we use first layer's model parameters as input to \attclf which outputs the training data's property.

\Hypar{Metrics}\label{sec:metrics} We describe different metrics to measure the effectiveness of inference based attestation. A model trained on a dataset with \preq is considered as the positive class. Accuracy indicates the success of \prover's model and \verifier's shadow models on the task. 
True Acceptance Rate (TAR) measures the fraction of models where \verifier correctly attests that \modelcheck was indeed trained from a dataset with \preq. True Rejection Rate (TRR) measures the fraction of models where \verifier correctly rejects attestation of \modelcheck w.r.t. \preq. 
FAR and {FRR} measure the extent of \verifier incorrectly accepting or rejecting attestation respectively.
Equal Error Rate (EER) indicates the value at which the FAR and FRR are equal. TAR and TRR should ideally be 1.00 while FAR, FRR and EER be 0.00.

For cryptographic attestation, we indicate computation cost ($\CostCryptComp$) as the execution time for \dc and secure model training; and communication cost ($\CostCryptComm$) as the amount of data transferred during attestation. 
For hybrid attestation 
with fixed FAR where \prover relies on cryptographic fallback, the expected cost is $\Pinf$ $\times$ $\CostInf$ + $\Pfallback$ $\times$ $\CostCrypt$ where $\Pinf=1$. As $\omega_{inf} \ll \omega_{crpt}$, the cost reduces to $\Pfallback$ $\times$ $\CostCrypt$. Similarly, for fixed FRR, \verifier conducts spot-checks with a probability of $\Pspotcheck$. The expected cost in this case is $\Pspotcheck \times \CostCrypt$. Both $\Pfallback$ and $\Pspotcheck$ are computed on \verifiertestdata.

\section{Experimental Evaluation}\label{sec:evaluation}

For different requirements: effectiveness, adversarial robustness and efficiency, we first evaluate the inference-based and cryptographic attestation and identify their limitations, then evaluate the hybrid attestation.

\subsection{Inference-based Attestation}\label{sec:inferenceeval}

\Hypar{Effectiveness (\ref{req2})} We first evaluate the effectiveness of \attclf in distinguishing between models trained on \proverdata with \preq and !\preq using AUC score under FAR-TAR curves \begin{submit}(see Appendix A of our full paper~\cite{duddu2023attesting}).\end{submit}
\begin{preprint}(see Appendix~\ref{app:windowsize}).\end{preprint} 
We find that for some \preq values, \attclf is less effective. Hence, we relax the attestation requirement to exactly match \preq by increasing the window size to $\pm$1 (i.e., classify between \{\preq-1,\preq,\preq+1\} and !\{\preq-1,\preq,\preq+1\})~\footnote{We  continue to use \preq and !\preq to refer to these windows.}. Based on the results, we identify the best window sizes on \verifiertestdata: $\pm$1 for all \preq for \boneage and \arxiv; 0 for the edge \preq values (i.e., ``0.00'' and ``1.00'') and $\pm$1 for all middle \preq values for \racecensus and \sexcensus. \verifier can make these decisions on \verifiertestdata before finalizing \attclf.






\makeparafit
Assuming that \verifier fixes (a) FAR, or (b) FRR at 5\%, we present the corresponding TRR and TAR in Table~\ref{tab:effectiveness} on \verificationdata. 
At either end of the spectrum of \preq values, attestation is effective (high TRR/TAR).
However, we observe a high FAR and FRR for the middle \preq values 
indicating that attestation is less effective. Furthermore, we provide EER values that, for specific \preq values, demonstrate lower rates than both FAR and FRR. This implies the existence of a more optimal threshold than the currently used $5\%$.
In summary, inference-based attestation is \textit{ineffective for certain \preq values} and cannot be used on its own.

\begin{table}[h]
\vspace{-0.6cm}
\setlength\tabcolsep{1pt}
\scriptsize
\caption{TAR and TRR with 5\% thresholds for FAR and FRR respectively along with EER across different \preq windows on \verificationdata. The \preq value within the window is indicated in \textbf{bold}. Edge \preq values have higher effectiveness than middle \preq values due to higher distinguishability in first layer parameters~\cite{suripropinf}.}
\centering
\begin{tabular}{cc} 

\begin{tabular}{c|c|c|c}
\toprule
\multicolumn{4}{c}{\arxiv} \\
\toprule
\textbf{\preq Range}  & \textbf{TAR}  & \textbf{TRR} & \textbf{EER} \\
 \toprule
\{\textbf{9}, 10\}  & 1.00 & 0.99 & 0.02 \\ 
\{9, \textbf{10}, 11\} & 1.00 & 1.00 & 0.01 \\ 
\{10, \textbf{11}, 12\} & 0.24 & 0.83 & 0.16\\ 
\{11, \textbf{12}, 13\} & 0.61 & 0.68 & 0.19\\ 
\{12, \textbf{13}, 14\} & 0.78 & 0.85 & 0.10 \\ 
\{13, \textbf{14}, 15\} & 0.92 & 0.93 & 0.07\\ 
\{14, \textbf{15}, 16\} & 0.87 & 0.90 & 0.08\\ 
\{15, \textbf{16}, 17\} & 1.00 & 1.00 & 0.00 \\ 
\{16, \textbf{17}\} & 1.00 & 1.00 & 0.00\\ 
 \bottomrule
\end{tabular}
\hspace{1cm}
& 
\begin{tabular}{c|c|c|c}
\toprule
\multicolumn{4}{c}{\boneage} \\
\toprule
\textbf{\preq Range} & \textbf{TAR}  & \textbf{TRR} & \textbf{EER} \\
 \toprule
\{\textbf{0.20}, 0.30\} & 0.96 & 0.96 & 0.03 \\ 
\{0.20, \textbf{0.30}, 0.40\} & 0.99 & 1.00 & 0.02\\ 
\{0.30, \textbf{0.40}, 0.50\} & 0.87 & 0.88 & 0.09\\ 
\{0.40, \textbf{0.50}, 0.60\} & 0.53 & 0.65 & 0.21\\ 
\{0.50, \textbf{0.60}, 0.70\} & 0.39 & 0.72 & 0.25\\ 
\{0.60, \textbf{0.70}, 0.80\} & 0.98 & 0.98 & 0.03\\ 
\{0.70, \textbf{0.80}\} & 0.95 & 0.95 & 0.05\\ 
 \bottomrule
\end{tabular} \\
\end{tabular}

\vspace{0.5cm} 

\begin{tabular}{cc} 

\begin{tabular}{c|c|c|c}
\toprule
\multicolumn{4}{c}{\sexcensus} \\
\toprule
\textbf{\preq Range} & \textbf{TAR}  & \textbf{TRR} & \textbf{EER} \\
 \toprule
 \{\textbf{0.00}\} & 1.00 & 1.00 & 0.00\\ 
\{0.00, \textbf{0.10}, 0.20\} & 0.49 & 0.49 & 0.19\\ 
 \{0.10, \textbf{0.20}, 0.30\} & 0.70 & 0.72 & 0.14\\ 
\{0.20, \textbf{0.30}, 0.40\} & 0.23 & 0.56 & 0.25\\ 
\{0.30, \textbf{0.40}, 0.50\} & 0.12 & 0.30 & 0.37\\ 
\{0.40, \textbf{0.50}, 0.60\} & 0.13 & 0.23 & 0.41\\ 
\{0.50, \textbf{0.60}, 0.70\} & 0.15 & 0.22 & 0.34\\
\{0.60, \textbf{0.70}, 0.80\} & 0.12 & 0.26 & 0.35\\ 
\{0.70, \textbf{0.80}, 0.90\} & 0.59 & 0.58 & 0.19\\ 
\{0.80, \textbf{0.90}, 1.00\} & 0.60 & 0.59 & 0.19\\ 
 \{\textbf{1.00}\} & 1.00 & 1.00 & 0.00\\ 
 \bottomrule
\end{tabular}
\hspace{1cm}
& 
\begin{tabular}{c|c|c|c}
\toprule
\multicolumn{4}{c}{\racecensus} \\
\toprule
\textbf{\preq Range} & \textbf{TAR}  & \textbf{TRR} & \textbf{EER} \\
 \toprule
 \{\textbf{0.00}\} & 1.00 & 1.00 & 0.00\\ 
 \{0.00, \textbf{0.10}, 0.20\} & 0.21 & 0.64 & 0.19 \\
 \{0.10, \textbf{0.20}, 0.30\} & 0.75 & 0.89 & 0.10\\ 
 \{0.20, \textbf{0.30}, 0.40\} & 0.22 & 0.59 & 0.23\\ 
 \{0.30, \textbf{0.40}, 0.50\} & 0.14 & 0.16 & 0.39\\ 
 \{0.40, \textbf{0.50}, 0.60\} & 0.10 & 0.15 & 0.42\\ 
 \{0.50, \textbf{0.60}, 0.70\} & 0.13 & 0.26 & 0.39\\
 \{0.60, \textbf{0.70}, 0.80\} & 0.05 & 0.36 & 0.32\\ 
 \{0.70, \textbf{0.80}, 0.90\} & 0.65 & 0.77 & 0.13\\ 
 \{0.80, \textbf{0.90}, 1.00\} & 0.35 & 0.41 & 0.26\\ 
 \{\textbf{1.00}\} & 1.00 & 1.00 & 0.00\\ 
 \bottomrule
\end{tabular} \\
\end{tabular}

\label{tab:effectiveness}
\end{table}

\Hypar{Robustness (\ref{req3})} \malprover can fool \verifier by modifying \modelcheck's first layer parameters (\firstlayer) to trigger FA. \malprover adds adversarial noise $\delta$ to the first layer parameters: \firstlayerdelta where
$\delta =argmax_{||\delta||_p<\epsilon}$ $L($ \attclf$($\firstlayerdelta$),$\preq~$)$, $L$ is the \attclf's loss and $||\cdot||_p$ is the $l_p$ norm to ensure $\delta$ to minimize accuracy degradation. Since, \malprover does not have access to \attclf, they train a ``substitute model'' on~\proverdata which mimics \attclf. For worst-case analysis under the attack, we assume that the substitute model's architecture is the same as \attclf. $\delta$ is then computed with respect to this substitute model and the FA are expected to transfer to \attclf. To restore any \modelcheck's accuracy loss, \malprover can freeze \firstlayer and fine-tune the remaining layers. We empirically evaluate this and confirm that accuracy of model after fine-tuning is close to the original accuracy while still being able to fool the attestation 
\begin{submit}(results in Appendix C of our full paper~\cite{duddu2023attesting}).\end{submit} 
\begin{preprint}(results in Appendix~\ref{app:accfinetuning}).\end{preprint} 
We use Autoattack~\cite{croce2020reliable} with $\epsilon = 8/255$, and $L_\infty$ norm for the distance function. As \malprover has access to the models shared with \verifier, we evaluate on \verificationdata.

\Itpar{Attack Success} \malprover wins if \attclf incorrectly classifies perturbed models as having been trained with \preq.
We measure the attack success using FAR. Note that the FAR here is restricted to \verificationdata containing models with adversarial noise. Under ``w/o Defence'' in Table~\ref{tab:robustness}, the high FAR values indicate that the attack is indeed successful (\attclf is not robust).

\begin{table}[!htb]
\setlength\tabcolsep{1pt}
\scriptsize
\centering
\caption{\textbf{Robustness against first layer parameter perturbations with and without a defense.} Utility ($\mathcal{U}$) is calculated using AUC on FAR-TAR for a clean dataset. FAR $\rightarrow$ lack of robustness. \colorbox{green!25}{Green} $\rightarrow$ $\mathcal{U}$ decreases or FAR $<5\%$, \colorbox{red!25}{red} $\rightarrow$  $\mathcal{U}$ increases or FAR $\geq5\%$.}

\begin{tabular}{c|c|c|c|c}
\toprule
\multicolumn{5}{c}{\textbf{\arxiv}} \\
\textbf{$p_{req}$ Range}  & \multicolumn{2}{c|}{w/o Defence}  & \multicolumn{2}{c}{w/ Defence} \\
& $\mathcal{U}$  &  FAR & $\mathcal{U}$ & FAR \\ 
\toprule
\{\textbf{9}, 10\}  & 1.00 $\pm$ 0.00 & 1.00 $\pm$ 0.00 & \cellcolor{green!30}1.00 $\pm$ 0.01 & \cellcolor{green!30}0.00 $\pm$ 0.00 \\ 
\{9, \textbf{10}, 11\} & 1.00 $\pm$ 0.01 & 1.00 $\pm$ 0.00  & \cellcolor{green!30}1.00 $\pm$ 0.00 & \cellcolor{green!30}0.00 $\pm$ 0.00 \\ 
\{10, \textbf{11}, 12\} & 0.92 $\pm$ 0.00 & 1.00 $\pm$ 0.00 & \cellcolor{green!30}0.88 $\pm$ 0.00 & \cellcolor{green!30}0.00 $\pm$ 0.00 \\ 
\{11, \textbf{12}, 13\} & 0.96 $\pm$ 0.00 & 1.00 $\pm$ 0.00 & \cellcolor{green!30}0.96 $\pm$ 0.00 & \cellcolor{green!30}0.00 $\pm$ 0.00  \\ 
\{12, \textbf{13}, 14\} & 0.93 $\pm$ 0.00 & 1.00 $\pm$ 0.00 & \cellcolor{green!30}0.96 $\pm$ 0.01 & \cellcolor{green!30}0.00 $\pm$ 0.00 \\ 
\{13, \textbf{14}, 15\} & 0.99 $\pm$ 0.00 & 1.00 $\pm$ 0.00 & \cellcolor{green!30}0.96 $\pm$ 0.01 & \cellcolor{green!30}0.00 $\pm$ 0.00 \\ 
\{14, \textbf{15}, 16\} & 0.99 $\pm$ 0.00 & 1.00 $\pm$ 0.00 & \cellcolor{green!30}1.00 $\pm$ 0.01 & \cellcolor{green!30}0.00 $\pm$ 0.00 \\ 
\{15, \textbf{16}, 17\} & 1.00 $\pm$ 0.00 & 1.00 $\pm$ 0.00 & \cellcolor{green!30}1.00 $\pm$ 0.00 & \cellcolor{green!30}0.00 $\pm$ 0.00\\
\{16, \textbf{17}\} & 1.00 $\pm$ 0.00 & 1.00 $\pm$ 0.00 & \cellcolor{green!30}1.00 $\pm$ 0.01 & \cellcolor{green!30}0.00 $\pm$ 0.00\\ 
 \bottomrule

\toprule
\multicolumn{5}{c}{\textbf{\boneage}} \\
\textbf{$p_{req}$ Range}  & \multicolumn{2}{c|}{w/o Defence}  & \multicolumn{2}{c}{w/ Defence} \\
& $\mathcal{U}$  &  FAR & $\mathcal{U}$ & FAR \\ 
\toprule
\{\textbf{0.20}, 0.30\} &  0.99 $\pm$ 0.00 & 1.00 $\pm$ 0.00 & \cellcolor{green!30}0.99 $\pm$ 0.00 & \cellcolor{green!30}0.00 $\pm$ 0.00 \\ 
\{0.20, \textbf{0.30}, 0.40\} & 0.99 $\pm$ 0.00 & 1.00 $\pm$ 0.00 & \cellcolor{green!30}0.99 $\pm$ 0.00 & \cellcolor{green!30}0.00 $\pm$ 0.00 \\ 
\{0.30, \textbf{0.40}, 0.50\} & 0.92 $\pm$ 0.00 & 1.00 $\pm$ 0.00 & \cellcolor{green!30}0.92 $\pm$ 0.00 & \cellcolor{green!30}0.00 $\pm$ 0.00 \\ 
\{0.40, \textbf{0.50}, 0.60\} & 0.86 $\pm$ 0.00 & 1.00 $\pm$ 0.00 & \cellcolor{green!30}0.84 $\pm$ 0.00 & \cellcolor{green!30}0.00 $\pm$ 0.01 \\ 
\{0.50, \textbf{0.60}, 0.70\} & 0.87 $\pm$ 0.00 & 0.28 $\pm$ 0.00  & \cellcolor{green!30}0.85 $\pm$ 0.00 & \cellcolor{red!30}0.25 $\pm$ 0.00 \\ 
\{0.60, \textbf{0.70}, 0.80\} & 0.99 $\pm$ 0.00 & 1.00 $\pm$ 0.00  & \cellcolor{green!30}0.99 $\pm$ 0.00 & \cellcolor{green!30}0.02 $\pm$ 0.00\\ 
\{0.70, \textbf{0.80}\} & 0.95 $\pm$ 0.00 & 0.04 $\pm$ 0.00 & \cellcolor{green!30}0.95 $\pm$ 0.00 & \cellcolor{green!30}0.00 $\pm$ 0.00 \\ 
 \bottomrule

\toprule
\multicolumn{5}{c}{\textbf{\sexcensus}} \\
\textbf{$p_{req}$ Range}  & \multicolumn{2}{c|}{w/o Defence}  & \multicolumn{2}{c}{w/ Defence} \\
& $\mathcal{U}$  &  FAR & $\mathcal{U}$ & FAR \\ 
\toprule
 \{\textbf{0.00}\} & 1.00 $\pm$ 0.00 & 0.00 $\pm$ 0.00 & \cellcolor{green!30}1.00 $\pm$ 0.00 & \cellcolor{green!30}0.00 $\pm$ 0.00 \\ 
 \{0.00, \textbf{0.10}, 0.20\} & 0.83 $\pm$ 0.01 & 0.26 $\pm$ 0.05  & \cellcolor{green!30}0.82 $\pm$ 0.01 & \cellcolor{green!30}0.01 $\pm$ 0.01 \\
 \{0.10, \textbf{0.20}, 0.30\} & 0.92 $\pm$ 0.00 & 0.12 $\pm$ 0.02 &  \cellcolor{green!30}0.92 $\pm$ 0.01 & \cellcolor{green!30}0.04 $\pm$ 0.01\\ 
 \{0.20, \textbf{0.30}, 0.40\} & 0.78 $\pm$ 0.01 & 0.11 $\pm$ 0.02 & \cellcolor{green!30}0.79 $\pm$ 0.00 & \cellcolor{red!30}0.10 $\pm$ 0.02 \\ 
 \{0.30, \textbf{0.40}, 0.50\} & 0.66 $\pm$ 0.00 & 0.34 $\pm$ 0.10 & \cellcolor{green!30}0.66 $\pm$ 0.01 & \cellcolor{red!30}0.29 $\pm$ 0.03 \\ 
 \{0.40, \textbf{0.50}, 0.60\} & 0.67 $\pm$ 0.01 & 0.39 $\pm$ 0.02 & \cellcolor{green!30}0.67 $\pm$ 0.00 & \cellcolor{red!30}0.22 $\pm$ 0.03 \\ 
 \{0.50, \textbf{0.60}, 0.70\} & 0.62 $\pm$ 0.01 & 0.19 $\pm$ 0.02 & \cellcolor{green!30}0.62 $\pm$ 0.01 & \cellcolor{red!30}0.14 $\pm$ 0.01 \\
 \{0.60, \textbf{0.70}, 0.80\} & 0.68 $\pm$ 0.00 &0.48 $\pm$ 0.01  & \cellcolor{green!30}0.68 $\pm$ 0.01 & \cellcolor{red!30}0.35 $\pm$ 0.04 \\ 
 \{0.70, \textbf{0.80}, 0.90\} & 0.89 $\pm$ 0.00 & 0.32 $\pm$ 0.02  & \cellcolor{green!30}0.89 $\pm$ 0.00 & \cellcolor{red!30}0.32 $\pm$ 0.05 \\ 
 \{0.80, \textbf{0.90}, 1.00\} &  0.89 $\pm$ 0.01 & 0.65 $\pm$ 0.03  & 
 \cellcolor{green!30} 0.89 $\pm$ 0.00 & \cellcolor{red!30} 0.22 $\pm$ 0.03 \\ 
 \{\textbf{1.00}\} &  1.00 $\pm$ 0.00 & 0.02 $\pm$ 0.00 & 
 \cellcolor{green!30} 1.00 $\pm$ 0.00 &  \cellcolor{green!30} 0.00 $\pm$ 0.00 \\ 
 \bottomrule

\toprule
\multicolumn{5}{c}{\textbf{\racecensus}} \\
\textbf{$p_{req}$ Range}  & \multicolumn{2}{c|}{w/o Defence}  & \multicolumn{2}{c}{w/ Defence} \\
& $\mathcal{U}$  &  FAR & $\mathcal{U}$ & FAR \\ 
\toprule
 \{\textbf{0.00}\} & 1.00 $\pm$ 0.00 & 0.00 $\pm$ 0.00 & \cellcolor{green!30}1.00 $\pm$ 0.00 & \cellcolor{green!30}0.00 $\pm$ 0.00 \\ 
  \{0.00, \textbf{0.10}, 0.20\} & 0.75 $\pm$ 0.01 & 0.82 $\pm$ 0.02 & \cellcolor{green!30}0.77 $\pm$ 0.01 & \cellcolor{red!30}0.35 $\pm$ 0.03 \\
 \{0.10, \textbf{0.20}, 0.30\} & 0.95 $\pm$ 0.00 & 0.60 $\pm$ 0.03 & \cellcolor{green!30}0.95 $\pm$ 0.00 & \cellcolor{green!30}0.08 $\pm$ 0.01 \\ 
 \{0.20, \textbf{0.30}, 0.40\} & 0.71 $\pm$ 0.01 & 0.83 $\pm$ 0.02 & \cellcolor{green!30}0.71 $\pm$ 0.01 & \cellcolor{red!30}0.13 $\pm$ 0.05\\ 
 \{0.30, \textbf{0.40}, 0.50\} & 0.75 $\pm$ 0.00 & 0.82 $\pm$ 0.07 & \cellcolor{green!30}0.74 $\pm$ 0.00 & \cellcolor{green!30}0.05 $\pm$ 0.01\\ 
 \{0.40, \textbf{0.50}, 0.60\} & 0.64 $\pm$ 0.00 & 0.79 $\pm$ 0.05 & \cellcolor{green!30}0.63 $\pm$ 0.00 & \cellcolor{red!30}0.11 $\pm$ 0.03 \\ 
 \{0.50, \textbf{0.60}, 0.70\} & 0.60 $\pm$ 0.01 & 0.31 $\pm$ 0.02  & \cellcolor{green!30}0.60 $\pm$ 0.00 & \cellcolor{red!30}0.31 $\pm$ 0.02 \\
\{0.60, \textbf{0.70}, 0.80\} & 0.83 $\pm$ 0.00 & 0.30 $\pm$ 0.02 & \cellcolor{green!30}0.82 $\pm$ 0.01 & \cellcolor{red!30}0.26 $\pm$ 0.01 \\ 
  \{0.70, \textbf{0.80}, 0.90\} & 0.96 $\pm$ 0.00 & 0.10 $\pm$ 0.01 &  \cellcolor{green!30} 0.96 $\pm$ 0.01 & 
 \cellcolor{red!30} 0.19 $\pm$ 0.00 \\ 
 \{0.80, \textbf{0.90}, 1.00\} &  0.70 $\pm$ 0.01 & 0.46 $\pm$ 0.03 & \cellcolor{green!30} 0.75 $\pm$ 0.01 & \cellcolor{red!30} 0.34 $\pm$ 0.02  \\ 
  \{\textbf{1.00}\} &  1.00 $\pm$ 0.00 & 0.00 $\pm$ 0.00 & \cellcolor{green!30} 1.00 $\pm$ 0.00 & \cellcolor{green!30} 0.00 $\pm$ 0.00  \\ 
 \bottomrule
\end{tabular}
\label{tab:robustness}
\end{table}

\Itpar{Improving Robustness} We propose adversarial training of \attclf where \verifier includes models with adversarial noise to train \attclf. Our goal is to reduce FAR on perturbed models while retaining utility on clean \verificationdata.

We present the results of adversarial training under ``w/ Defence'' in Table~\ref{tab:robustness}. First, the FAR values in ``w/ Defence'' are lower (than in ``w/o Defence''). Hence, adversarial training of \attclf successfully mitigates the perturbation attack, thus making inference-based attestation adversarially robust, satisfying \ref{req3}. Second, the difference in utility on clean \verificationdata (measured using AUC score under FAR-TAR curves indicated by $\mathcal{U}$) between ``w/ Defence'' and ``w/o Defence'' is small. We also evaluate the effectiveness using TRR, TAR, and EER on clean \verificationdata\xspace 
\begin{submit}
which are available in Appendix B of our full paper~\cite{duddu2023attesting}.
\end{submit}
\begin{preprint}
(see Appendix~\ref{app:robustVerdata}).
\end{preprint}
We use robust \attclf in the rest of the paper. Similar to \attclf, robust \attclf is still ineffective for some \preq.

\noindent\textbf{Efficiency (\ref{req4}).} \verifier trains multiple shadow models and \attclf. We measure the total training time to train 10 attestation classifiers and 1000 shadow models on a single NVIDIA A100 GPU. Training \attclf took a total of 200 mins for \boneage; 12 mins for \arxiv, ~6 mins for \sexcensus and \racecensus. Training 1000 shadow models took a total of 173 mins for \boneage, 123 mins for \arxiv and 50 mins for \sexcensus and \racecensus. 

\boneage, being an image dataset trained on a large neural network, takes the maximum time for both \attclf and shadow models. On the other hand, \census has a small number of tabular data records and with a small MLP classifier, takes the least training time. Note that this is a one-time cost which can be parallelized among multiple GPUs. Hence, \verifier's cost for inference-based attestation is reasonable. \attclf can then be used for multiple attestation for the same property and  has to be trained on new property only if $p_{req}$ changes.

\Hypar{Summary} Inference-based attestation satisfies \ref{req3} robustness and \ref{req4} efficiency but has a \textit{poor effectiveness}. 

\subsection{Cryptographic Attestation}\label{sec:cryptoeval}

\Hypar{Effectiveness (\ref{req2})} Cryptographic attestation operates over \proverdata confidentially to correctly check whether the distributional properties match \preq. Hence, we have \textit{zero FAR and FRR}.

\Hypar{Robustness (\ref{req3})} Using outsourcing, \prover's inputs are secret-shared between \serverone and \servertwo who learn nothing (non-colluding assumption) and have no incentive to cheat (semi-honest assumption). Furthermore, \prover \textit{only} performs the input sharing of the training data \proverdata and initial model weights, thus \prover cannot cheat during proof generation. Hence, this attestation is robust.

\Hypar{Efficiency (\ref{req4})} We use protocols for semi-honest parties. We present the computation and communication cost for a single cryptographic attestation in Table~\ref{tab:cryptocost}. We indicate the costs for \boneage and \census but omit the evaluation on \arxiv as there are no PyTorch frameworks for secure GNN training which is required for the CrypTen library. We observe that the cost for \dc is low, but the cost for secure ML training is high. Hence, cryptographic attestation is difficult to be used in practice for multiple \prover{s}.
\vspace{-1.0cm}
\begin{table}[!htb]
\caption{Computation ($\CostCryptComp$) and communication costs ($\CostCryptComm$) of cryptographic attestation for a single \prover averaged over 20 runs.}
\centering
\setlength\tabcolsep{4pt}
\renewcommand{\arraystretch}{1.2}
\scriptsize
\begin{tabular}{@{} l c c c c @{}}
\toprule
\textbf{Datasets} & \multicolumn{2}{c}{$\CostCryptComp$ (s)} & \multicolumn{2}{c}{$\CostCryptComm$ (GB)} \\ 
\cmidrule(lr){2-3} \cmidrule(lr){4-5}
 & DistCheck & Training & DistCheck & Training \\ 
\midrule
\textbf{\boneage} & 1.30 $\pm$ 0.05 & 1367.31 $\pm$ 27.95 & 0.01  & 228.54 \\  
\textbf{\racecensus} & 1.54 $\pm$ 0.15 & 1081.00 $\pm$ 17.00  & 0.01 & 874.06 \\ 
\textbf{\sexcensus} & 1.68 $\pm$ 0.15 & 2109.78 $\pm$ 65.20  & 0.01  & 1438.38 \\ 
\bottomrule
\end{tabular}
\label{tab:cryptocost}
\end{table}


\Hypar{Summary} Cryptographic attestation satisfies \ref{req2} effectiveness and \ref{req3} robustness, but \textit{lacks efficiency} which limits its scalability to multiple \prover{s}.

\subsection{Hybrid Attestation}\label{sec:hybrideval}

We present the effectiveness of hybrid attestation with a fixed FAR (or FRR), its impact on the respective FRR (or FAR) and the expected cost incurred.

\subsubsection{Fixed FAR Analysis} \label{sec:fixFAR}
Recall that on fixing FAR, rejected \prover{s} can request re-evaluation using cryptographic attestation as a fallback.

\begin{figure}[!htb]
\begin{minipage}[h]{0.49\linewidth}
\begin{center}
\includegraphics[width=0.8\linewidth]{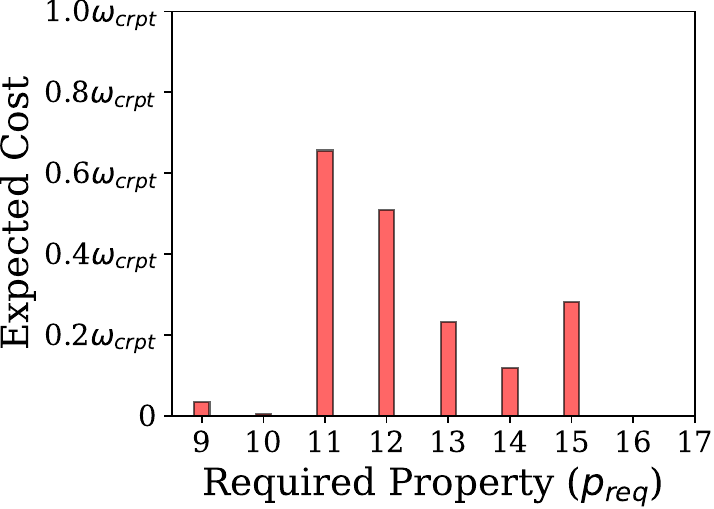}
\subcaption{\arxiv}
\end{center} 
\end{minipage}
\hfill
\begin{minipage}[h]{0.49\linewidth}
\begin{center}
\includegraphics[width=0.8\linewidth]{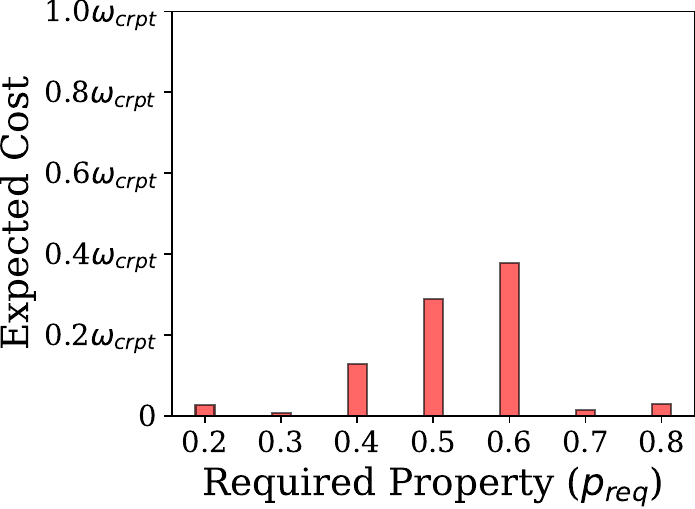} 
\subcaption{\boneage}
\end{center}
\end{minipage}
\vfill
\begin{minipage}[h]{0.49\linewidth}
\begin{center}
\includegraphics[width=0.8\linewidth]{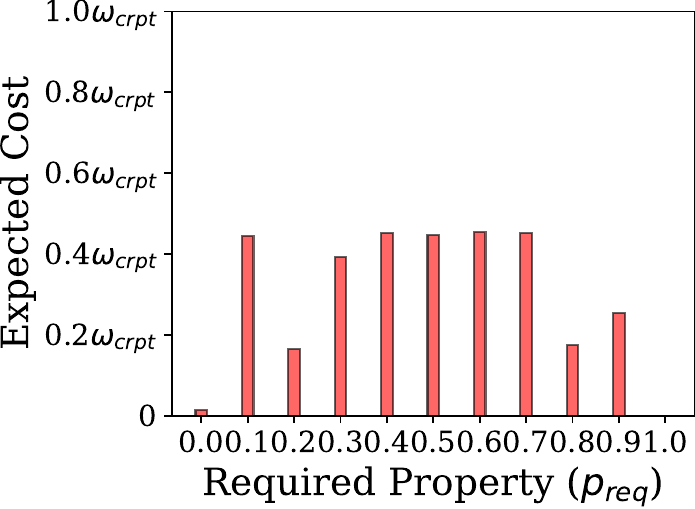} 
\subcaption{\racecensus}
\end{center}
\end{minipage}
\hfill
\begin{minipage}[h]{0.49\linewidth}
\begin{center}
\includegraphics[width=0.8\linewidth]{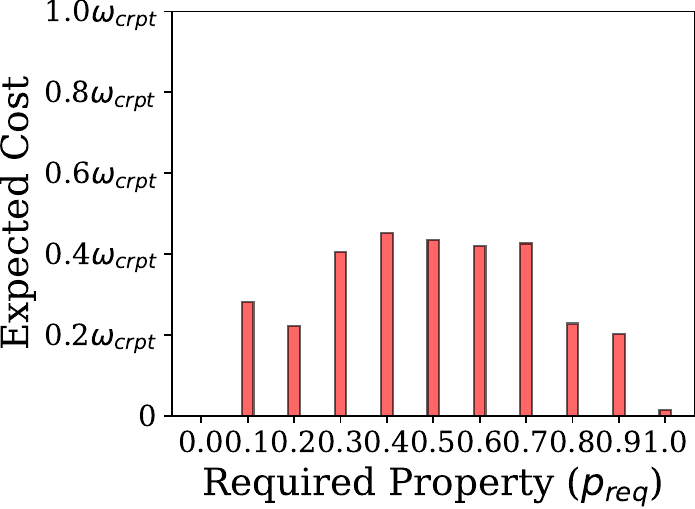} 
\subcaption{\sexcensus}
\end{center}
\end{minipage}
\caption{\textbf{Fixed FAR@5\%: Expected cost on \verificationdata.} $\CostCrypt$ is a placeholder for $\CostCryptComp$ and $\CostCryptComm$. Expected cost is less than cryptographic attestation (=$\CostCrypt$).}
\label{fig:worstcase}

\vspace{-0.6cm}
\end{figure}

\Hypar{Effectiveness (\ref{req2})} Hybrid attestation will \textit{not change the 5\% fixed FAR value}. We compute the effective FRR on using cryptographic attestation as a fallback. In practice, only \prover{s} with FR have the incentive to request re-evaluation using cryptographic attestation. If such \prover{s} undergo re-evaluation, \textit{FRR is zero} as the cryptographic attestation will rectify any erroneous decision.

\Hypar{Expected Cost (\ref{req4})} We evaluate the expected cost on \verificationdata. This gives the actual estimate of the cost incurred during attestation. We assume that \prover{s} are rational, so only \prover{s} with FR will request a re-evaluation using cryptographic attestation. Here, to compute the expected cost, $\Pfallback = \frac{\Nrejects}{\Ntotal}$, where $\Nrejects$ is the total number of rejected \prover{s}.

We present the expected cost in Figure~\ref{fig:worstcase} where the values of $\CostCrypt$, a placeholder for $\CostCryptComp$ or $\CostCryptComm$, are from Table~\ref{tab:cryptocost}. Compared to cryptographic attestation with an expected cost of $\CostCrypt$, hybrid attestation has a lower expected cost across different datasets and \preq. Additionally, since $\Pfallback$ depends on $\Nrejects$ computed from inference-based attestation, the edge \preq values, where inference-based attestation is effective, have lower expected cost than middle \preq values.

\subsubsection{Fixing FRR} \label{sec:fixFRR} 
Here, recall that \verifier conducts random spot-checks to reduce FAR. 

\Hypar{Effectiveness (\ref{req2})} \verifier's choice of \Nspotcheck determines FAR. No spot-checks corresponds to same FAR as inference-based attestation while spot-checks for all accepted \prover{s} indicates zero FAR.

\Hypar{Efficiency (\ref{req4})} The expected cost incurred per \prover increases with \Nspotcheck. No spot-checks corresponds to no expected cryptographic cost while spot-checks for all accepted \prover{s} incurs a high expected cost. Hence, \verifier decides \Nspotcheck based on their application's requirement.

\begin{figure}[!htb]
\begin{minipage}[h]{0.49\linewidth}
\begin{center}
\includegraphics[width=0.8\linewidth]{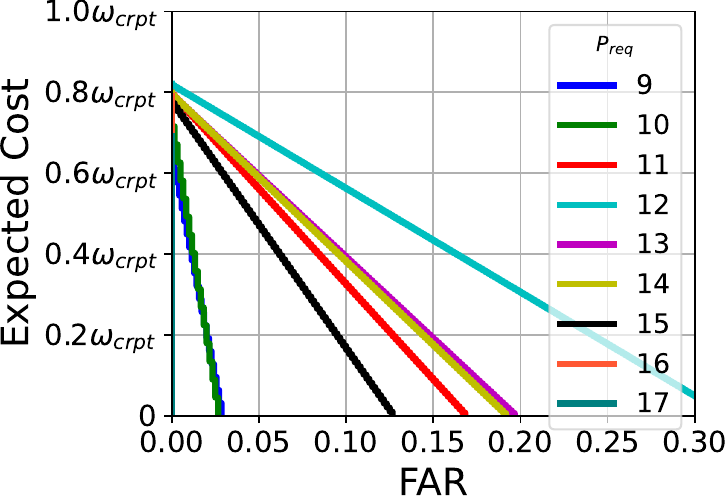} 
\subcaption{\arxiv}
\label{qwe5}
\end{center} 
\end{minipage}
\hfill
\begin{minipage}[h]{0.49\linewidth}
\begin{center}
\includegraphics[width=0.8\linewidth]{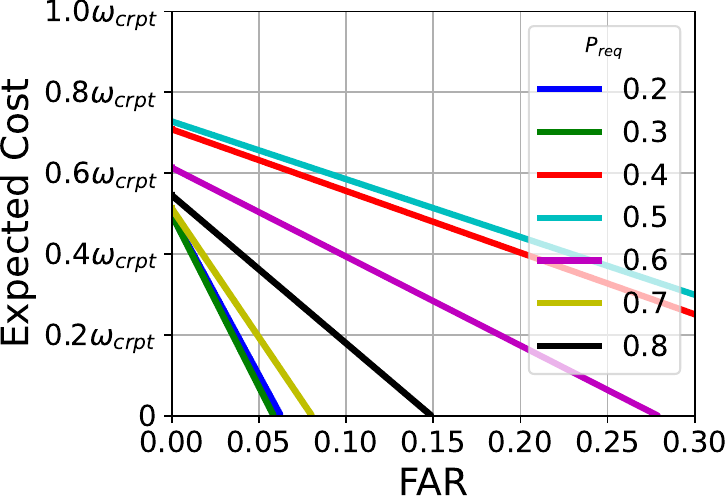} 
\subcaption{\boneage}
\label{qwe6}
\end{center}
\end{minipage}
\vfill
\begin{minipage}[h]{0.49\linewidth}
\begin{center}
\includegraphics[width=0.8\linewidth]{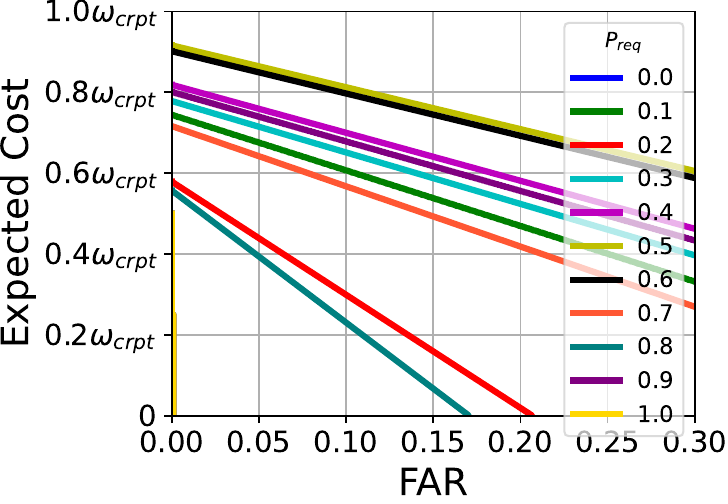} 
\subcaption{\racecensus}
\label{qwe7}
\end{center}
\end{minipage}
\hfill
\begin{minipage}[h]{0.49\linewidth}
\begin{center}
\includegraphics[width=0.8\linewidth]{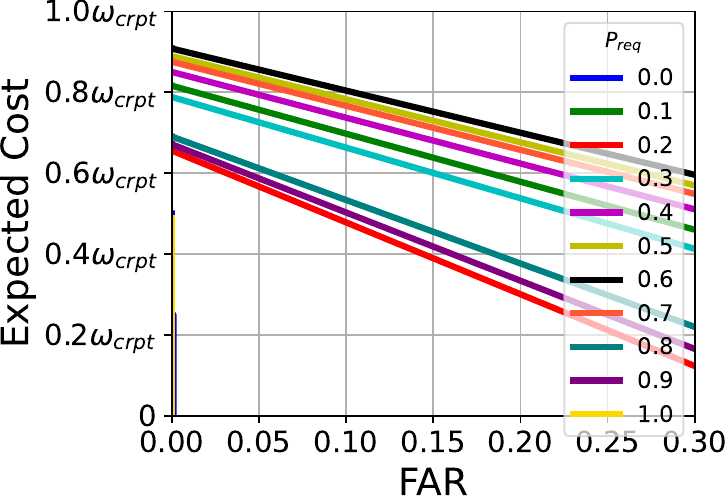} 
\subcaption{\sexcensus}
\label{qwe8}
\end{center}
\end{minipage}
\caption{\textbf{Fixed FRR@5\%: Trade-off between FAR and expected cost on varying \Nspotcheck on \verifiertestdata.} Expected cost is less than cryptographic attestation (=$\CostCrypt$) and effectiveness is better than inference-based attestation. $\CostCrypt$ is a placeholder for both $\CostCryptComp$ and $\CostCryptComm$.}
\label{fig:farVcostTest}
\vspace{-0.6cm}
\end{figure}

We present this trade-off between FAR and expected cost using cryptographic attestation by varying \Nspotcheck on \verifiertestdata in Figure~\ref{fig:farVcostTest}. We use $\Pspotcheck=\frac{\mathcal{N}_{spchk}}{\Ntotal}$ for the expected cost. Increasing \Nspotcheck, increases the expected cost while FAR decreases. Using Figure~\ref{fig:farVcostTest}, \verifier can determine \Nspotcheck. Once \verifier decides on a suitable \Nspotcheck using \verifiertestdata, the actual cost and FAR value can be read from a plot (similar to Figure~\ref{fig:farVcostTest}) for \verificationdata corresponding to the chosen \Nspotcheck 
\begin{submit}
(see Appendix E of our full paper~\cite{duddu2023attesting}).
\end{submit}
\begin{preprint}
(see Appendix~\ref{sec:hybridverdata}).  
\end{preprint}

Carefully choosing \Nspotcheck leads to a notable reduction in FAR compared with \Nspotcheck= 0 (x-axis). Additionally, we have lower expected cost compared to conducting spot-checks for all accepted \prover{s} and purely cryptographic attestation (y-axis where \Nspotcheck= 0).

\Hypar{Summary} Hybrid attestation is more effective than inference-based attestation and incurs a lower expected cost than cryptographic attestation.
\section{Related Work}\label{sec:related}

\noindent\textbf{Property Attestation} in trusted computing~\cite{propAtt,DBLP:conf/trust/KostiainenAE11} allows attesting if \prover's system satisfies the desired (security) requirements without revealing its specific software or hardware configuration. We are the first to introduce such a notion in ML while presenting mechanisms for distributional property attestation.

\noindent\textbf{Property Testing} compares the closeness of two distributions using mean and standard deviation~\cite{Canonne2020ASO}. In contrast, we need \verifier to test if \proverdata corresponds to the distribution expected by \verifier \emph{without} having access to \proverdata. One can conceivably implement property testing using 2PC, which will be similar to our cryptographic attestation protocol.
Chang et al.~\cite{287314} combine MPC and ZKP with property testing to check for data quality. However, they consider a different setting with multiple parties and their evaluation does not account for ML.

\noindent\textbf{Auditing ML Models} has been explored by adapting membership inference attacks to check for compliance with ``Right to Erasure''~\cite{audioAuditor,auditingText,unlearnaudit1,unlearnaudit2}. 
Juarez et al.~\cite{juarez2022black} use property inference to check for a specific case of distribution shift from balanced data (\preq = $0.5$). Our scheme is broader by allowing attestations for arbitrary properties as required by \verifier. Further, their scheme is insufficient for attestation as it lacks effectiveness (\ref{req2}) and robustness (\ref{req3}). We address these concerns in our work.
Additionally, cryptographic primitives can help audit models for fairness w.r.t. output predictions~\cite{blindjustice,auditfair1,auditfair2} which is different from property attestation considered in this work. 
``Proof-of-Learning'' (also known as proof-of-training) proves that a model was trained on a specific dataset using ML~\cite{jia2021proof}. However, such ML based schemes can be evaded~\cite{zhang2022adversarial,fang2023proofoflearning}.
Garg et al.~\cite{zkPoT} propose proof-of-training by combining ZKP with MPC-in-the-head in a concurrent and independent work. They mention the possibility of attesting properties, but do not implement it. Moreover, their approach is limited to logistic regression (e.g.,~\cite{eisenhofer2022verifiable}). Our hybrid approach using MPC is currently the best available approach that scales to larger models.

\noindent\textbf{Property Inference Attacks} have been explored in different domains: image, graphs and tabular data, threat models and classification tasks~\cite{zhangpropinf,ganpropinf,suripropinf,ganjupropinf,ateniesepropinf,fedLearningAttInf,suri2022dissecting,snap,chase2021property,ganpropinf}. Defending against them is an open problem~\cite{distprivacy,chen2022protecting,hartmann2022distribution,suri2022dissecting}.

\noindent\textbf{Privacy-Preserving ML} is an active research field, with much focus placed on cryptographic methods for privacy-preserving supervised and deep learning inference and training~\cite{mohassel2017secureml,USENIX:PSSY21ABY2.0,icml:Keller022,ccs:MohasselR18}. See the survey~\cite{sokSecureInf} for an overview.

\section{Discussions}\label{sec:discussions}

\noindent\textbf{Outsourcing as a Trade-off between Security and Efficiency.}
We cannot run cryptographic attestation using semi-honest 2PC protocols directly between a malicious prover~\malprover and \verifier, because \malprover can easily change the outcome of ``DistCheck'' for \preq by flipping its share of the output bit.

\Itpar{Proof} Let $[\cdot]^1$ denote shares held by \malprover and $[\cdot]^2$ shares held by \verifier. Let $\mathcal{D}_{\mathcal{P}{mal}}$ denote the dataset that only \malprover holds and wants to use for cryptographic attestation. Furthermore, let $v$ denote the true result of the verification and $out$ the output of the verification protocol. Since \preq is known to \malprover, it knows whether or not $\mathcal{D}_{\mathcal{P}{mal}}$ fulfills the requirement and can fool the \verifier by flipping the outcome, if the requirement is not met. If \malprover wants to flip the outcome of the verification, \malprover sets $[out]^1 = 1 \oplus [v]^1$, s.t. verification yields $out = [out]^1 \oplus [out]^2 = 1 \oplus [v]^1 \oplus [v]^2 = 1 \oplus v$, where $1 \oplus v = true$ iff the true outcome $v = false$. $\qed$
Alternatively, robustness against \malprover can be achieved with malicious protocols~\cite{zheng2021cerebroMalicious,Keller2020MPSPDZAV}, however, at a high cost. 
Then, how can we account for \malprover without using maliciously secure 2PC protocols? 
For this, we use secure outsourcing by introducing additional non-colluding semi-honest servers \serverone and \servertwo that carry out the cryptographic protocol on behalf of \prover and \verifier~\cite{Carter2015OutsourcingST}.

\Hypar{Alternative Protocols and their Limitations} Instead of outsourcing, we can replace MPC with other cryptographic protocols like ZKP. Non-Interactive ZKPs can be reused, thus the cost amortizes over multiple parties. 
However, they incur a high cost making them impractical as do other cryptographic mechanisms~\cite{sokSecureInf,zkPoT}. 
On the other hand, our MPC based approach can scale to neural networks. Further, TEEs offer an alternative approach, but may pose a deployment hurdle by requiring all \prover{s} and \verifier to have a TEE. Hence, designing more efficient protocols for property attestation is left as future work.

\Hypar{Relation with Fairness} Fairness involves a subsequent evaluation of model predictions to gauge the consistency of metrics like the false positive rate among various subgroups. The selection of an appropriate reference dataset holds significant importance, making it unclear whether the model is actually fairness~\cite{shamsabadi2023confidentialprofitt}. Biased datasets tend to yield more inequitable models compared to unbiased counterparts~\cite{preprocessing}. Hence, distribution equity is a prerequisite for fairness.

\Hypar{Whitebox Access and Inference Attacks by \verifier} Our setting is attestation for regulatory compliance, i.e., both \prover and \verifier \emph{co-operate} as both want attestation to succeed. If \verifier is a potential buyer, whitebox access to the model is natural. If \verifier is a regulator, whitebox access is still reasonable because \verifier is ``honest-but-curious'', i.e., \verifier may misuse any available information, but will not deviate from the specified protocol. Hence, \verifier \emph{must not be given the training dataset} (\proverdata), but can be trusted not to mount other inference attacks. Also, \textit{whitebox access is not needed} for cryptographic attestation: \verifier never sees \proverdata in the clear because the computation is over encrypted (secret-shared) data. 
Further, we can also add DP to minimize privacy risks without losing attestation accuracy as distribution inference is \emph{more successful} with DP \textit{assuming} \verifier knows the DP hyperparameters, which is reasonable in the attestation setting~\cite{suri2022dissecting}. Thus, we expect inference-based attestation will be \textit{more effective} with DP.

\begin{preprint}
\noindent\textbf{Malicious \verifier.}
For cryptographic attestation, a malicious \verifier does not get to interact with the protocol at all, but only receives the output shares from the outsourcing servers. Hence, they cannot manipulate the protocol to trigger a false rejection. For inference-based attestation, a malicious \verifier could indeed use an attestation classifier for a different property. In our setting of regulatory compliance, it is reasonable to assume that the \verifier is honest-but-curious and will not deviate from the specified protocol.
If it is necessary to accommodate a malicious \verifier, it can be done in a number of possible ways, each with its own cons: 
\begin{enumerate*}[label=\roman*),itemjoin={,\xspace}]
\item relying solely on cryptographic attestation (expensive) 
\item relying on the use of a hardware-assisted trusted execution environment (deployment requires additional assumption)
\item training \modelcheck with differential privacy (decrease in \modelcheck utility).
\end{enumerate*}
\end{preprint}
\section*{Acknowledgements}\label{sec:acknowledgements}
This work is supported in part by Intel (Private AI consortium) and the Government of Ontario.
Additionally, this project received funding from the European Research Council~(ERC) under the~European Union's Horizon~2020 research and innovation program~(grant agreement No.~850990 PSOTI). It was co-funded by the Deutsche Forschungsgemeinschaft~(DFG) within SFB~1119 CROSSING/236615297 and GRK~2050 Privacy \& Trust/251805230. 

\bibliographystyle{splncs04}
\bibliography{paper}

\begin{thebibliography}{10}
\providecommand{\url}[1]{\texttt{#1}}
\providecommand{\urlprefix}{URL }
\providecommand{\doi}[1]{https://doi.org/#1}

\bibitem{CTRSA:AtapoorSA22}
Atapoor, S., Smart, N.P., Alaoui, Y.T.: {Private Liquidity Matching Using
  {MPC}}. In: {CT-RSA} (2022)

\bibitem{ateniesepropinf}
Ateniese, G., Mancini, L.V., Spognardi, A., Villani, A., Vitali, D., Felici,
  G.: Hacking smart machines with smarter ones: How to extract meaningful data
  from machine learning classifiers. Int. J. Secur. Netw.  (2015)

\bibitem{Canonne2020ASO}
Canonne, C.L.: A survey on distribution testing: Your data is big. but is it
  blue? Theory of Computing  (2020)

\bibitem{Carter2015OutsourcingST}
Carter, H., Mood, B., Traynor, P., Butler, K.R.B.: Outsourcing secure two-party
  computation as a black box. In: Secur. Commun. Networks (2015)

\bibitem{287314}
Chang, I., Sotiraki, K., Chen, W., Kantarcioglu, M., Popa, R.: {HOLMES}:
  {E}fficient distribution testing for secure collaborative learning. In:
  USENIX Security (2023)

\bibitem{chase2021property}
Chase, M., Ghosh, E., Mahloujifar, S.: Property inference from poisoning.
  arXiv:2101.11073  (2021)

\bibitem{snap}
Chaudhari, H., Abascal, J., Oprea, A., Jagielski, M., Tramèr, F., Ullman, J.:
  {SNAP}: Efficient extraction of private properties with poisoning. In: S\&P
  (2023)

\bibitem{chen2022protecting}
Chen, M., Ohrimenko, O.: Protecting global properties of datasets with
  distribution privacy mechanisms. arXiv:2207.08367  (2022)

\bibitem{congress}
Congress, U.: H.r.6580 - algorithmic accountability act of 2022 (2022),
  \url{https://www.congress.gov/bill/117th-congress/house-bill/6580/text}

\bibitem{10136159}
Coston, A., Kawakami, A., Zhu, H., Holstein, K., Heidari, H.: A validity
  perspective on evaluating the justified use of data-driven decision-making
  algorithms. In: SaTML (2023)

\bibitem{croce2020reliable}
Croce, F., Hein, M.: Reliable evaluation of adversarial robustness with an
  ensemble of diverse parameter-free attacks. In: ICML (2020)

\bibitem{eur}
EC, E.C.: Regulation of the european parliament and of the council laying down
  harmonized rules on artificial intelligence (artificial intelligence act)
  (2021)

\bibitem{eisenhofer2022verifiable}
Eisenhofer, T., Riepel, D., Chandrasekaran, V., Ghosh, E., Ohrimenko, O.,
  Papernot, N.: Verifiable and provably secure machine unlearning.
  arXiv:2210.09126  (2022)

\bibitem{mozillaDeployment}
Englehardt, S.: Next steps in privacy-preserving telemetry with {P}rio.
  \url{https://blog.mozilla.org/security/2019/06/06/next-steps-in-privacy-preserving-telemetry-with-prio/}
  (2019)

\bibitem{fang2023proofoflearning}
Fang, C., Jia, H., Thudi, A., Yaghini, M., Choquette-Choo, C.A., Dullerud, N.,
  Chandrasekaran, V., Papernot, N.: Proof-of-learning is currently more broken
  than you think. arXiv:2208.03567  (2023)

\bibitem{ganjupropinf}
Ganju, K., Wang, Q., Yang, W., Gunter, C.A., Borisov, N.: Property inference
  attacks on fully connected neural networks using permutation invariant
  representations. In: CCS (2018)

\bibitem{zkPoT}
Garg, S., Goel, A., Jha, S., Mahloujifar, S., Mahmoody, M., Policharla, G.V.,
  Wang, M.: Experimenting with zero-knowledge proofs of training. In: CCS
  (2023)

\bibitem{GMW87}
Goldreich, O., Micali, S., Wigderson, A.: How to play any mental game. In: STOC
  (1987)

\bibitem{hartmann2022distribution}
Hartmann, V., Meynent, L., Peyrard, M., Dimitriadis, D., Tople, S., West, R.:
  Distribution inference risks: Identifying and mitigating sources of leakage.
  arXiv:2209.08541  (2022)

\bibitem{densenet}
Huang, G., Liu, Z., Van Der~Maaten, L., Weinberger, K.Q.: Densely connected
  convolutional networks. In: Computer Vision and Pattern Recognition (2017)

\bibitem{unlearnaudit2}
Huang, Y., Li, X., Li, K.: Ema: Auditing data removal from trained models. In:
  Medical Image Computing and Computer Assisted Intervention (2021)

\bibitem{jia2021proof}
Jia, H., Yaghini, M., Choquette-Choo, C.A., Dullerud, N., Thudi, A.,
  Chandrasekaran, V., Papernot, N.: Proof-of-learning: Definitions and
  practice. In: S\&P (2021)

\bibitem{juarez2022black}
Juarez, M., Yeom, S., Fredrikson, M.: Black-box audits for group distribution
  shifts. arXiv:2209.03620  (2022)

\bibitem{preprocessing}
Kamiran, F., Calders, T.: Data pre-processing techniques for classification
  without discrimination. Knowledge and Information Systems  (2011)

\bibitem{berkeleyDeployments}
Kaviani, D., Popa, R.A.: {M}{P}{C} {D}eployments.
  \url{https://mpc.cs.berkeley.edu} (2023)

\bibitem{distprivacy}
Kawamoto, Y., Murakami, T.: Local obfuscation mechanisms for hiding probability
  distributions. In: ESORICS (2019)

\bibitem{Keller2020MPSPDZAV}
Keller, M.: {MP-SPDZ}: A versatile framework for multi-party computation. CCS
  (2020)

\bibitem{icml:Keller022}
Keller, M., Sun, K.: {Secure Quantized Training for Deep Learning}. In: ICML
  (2022)

\bibitem{blindjustice}
Kilbertus, N., Gascon, A., Kusner, M., Veale, M., Gummadi, K., Weller, A.:
  Blind justice: Fairness with encrypted sensitive attributes. In: ICML (2018)

\bibitem{knott2021crypten}
Knott, B., et~al.: {CrypTen}: Secure multi-party computation meets machine
  learning. In: NeurIPS (2021)

\bibitem{DBLP:conf/trust/KostiainenAE11}
Kostiainen, K., Asokan, N., Ekberg, J.: Practical property-based attestation on
  mobile devices. In: TRUST (2011)

\bibitem{lindell2020secure}
Lindell, Y.: Secure multiparty computation. CACM  (2020)

\bibitem{unlearnaudit1}
Liu, X., Tsaftaris, S.A.: Have you forgotten? a method to assess if machine
  learning models have forgotten data. arXiv:2004.10129  (2020)

\bibitem{fedLearningAttInf}
Melis, L., Song, C., Cristofaro, E.D., Shmatikov, V.: Exploiting unintended
  feature leakage in collaborative learning. In: S\&P (2019)

\bibitem{audioAuditor}
Miao, Y., Xue, M., Chen, C., Pan, L., Zhang, J., Zhao, B.Z.H., Kaafar, D.,
  Xiang, Y.: The audio auditor: User-level membership inference in internet of
  things voice services. In: PETS (2021)

\bibitem{ccs:MohasselR18}
Mohassel, P., Rindal, P.: {ABY\({}^{\mbox{3}}\): {A} Mixed Protocol Framework
  for Machine Learning}. In: CCS (2018)

\bibitem{mohassel2017secureml}
Mohassel, P., Zhang, Y.: {SecureML}: A system for scalable privacy-preserving
  machine learning. In: S\&P (2017)

\bibitem{mpcallianceAlliance}
MPC-Alliance: {M}{P}{C} {A}lliance. \url{https://www.mpcalliance.org} (2023)

\bibitem{sokSecureInf}
Ng, L.L., Chow, S.M.: {SoK}: Cryptographic neural-network computation. In: S\&P
  (2023)

\bibitem{SoKMLPrivSec}
Papernot, N., McDaniel, P., Sinha, A., Wellman, M.P.: {SoK}: Security and
  privacy in machine learning. In: EuroS\&P (2018)

\bibitem{auditfair1}
Park, S., Kim, S., Lim, Y.s.: Fairness audit of machine learning models with
  confidential computing. In: WWW (2022)

\bibitem{PasquiniSplitInf}
Pasquini, D., Ateniese, G., Bernaschi, M.: Unleashing the tiger: Inference
  attacks on split learning. In: CCS (2021)

\bibitem{USENIX:PSSY21ABY2.0}
Patra, A., Schneider, T., Suresh, A., Yalame, H.: {ABY2.0: Improved
  Mixed-Protocol Secure Two-Party Computation}. In: {USENIX Security} (2021)

\bibitem{Riazi2018ChameleonAH}
Riazi, M.S., Weinert, C., Tkachenko, O., Songhori, E.M., Schneider, T.,
  Koushanfar, F.: Chameleon: A hybrid secure computation framework for machine
  learning applications. In: ASIACCS (2018)

\bibitem{propAtt}
Sadeghi, A.R., St\"{u}ble, C.: Property-based attestation for computing
  platforms: Caring about properties, not mechanisms. In: Workshop on New
  Security Paradigms (2004)

\bibitem{auditfair2}
Segal, S., Adi, Y., Pinkas, B., Baum, C., Ganesh, C., Keshet, J.: Fairness in
  the eyes of the data: Certifying machine-learning models. In: AIES (2021)

\bibitem{shamsabadi2023confidentialprofitt}
Shamsabadi, A.S., Wyllie, S.C., Franzese, N., Dullerud, N., Gambs, S.,
  Papernot, N., Wang, X., Weller, A.: Confidential-{PROFITT}: Confidential
  {PRO}of of fair training of trees. In: International Conference on Learning
  Representations (2023)

\bibitem{auditingText}
Song, C., Shmatikov, V.: Auditing data provenance in text-generation models.
  In: KDD (2019)

\bibitem{suripropinf}
Suri, A., Evans, D.: Formalizing and estimating distribution inference risks.
  PETS  (2022)

\bibitem{suri2022dissecting}
Suri, A., Lu, Y., Chen, Y., Evans, D.: Dissecting distribution inference. In:
  SaTML (2023)

\bibitem{mystique}
Weng, C., Yang, K., Xie, X., Katz, J., Wang, X.: Mystique: Efficient
  conversions for zero-knowledge proofs with applications to machine learning.
  In: USENIX Security (2021)

\bibitem{deepsets}
Zaheer, M., Kottur, S., Ravanbakhsh, S., Poczos, B., Salakhutdinov, R.R.,
  Smola, A.J.: Deep sets. In: NeurIPS (2017)

\bibitem{zhang2022adversarial}
Zhang, R., Liu, J., Ding, Y., Wang, Z., Wu, Q., Ren, K.: “adversarial
  examples” for proof-of-learning. In: S\&P (2022)

\bibitem{zhangpropinf}
Zhang, W., Tople, S., Ohrimenko, O.: Leakage of dataset properties in
  {Multi-Party} machine learning. In: USENIX Security (2021)

\bibitem{infAttGraphs}
Zhang, Z., Chen, M., Backes, M., Shen, Y., Zhang, Y.: Inference attacks against
  graph neural networks. In: USENIX Security (2022)

\bibitem{zheng2021cerebroMalicious}
Zheng, W., Deng, R., Chen, W., Popa, R.A., Panda, A., Stoica, I.: Cerebro: A
  platform for multi-party cryptographic collaborative learning. In: USENIX
  Security (2021)

\bibitem{ganpropinf}
Zhou, J., Chen, Y., Shen, C., Zhang, Y.: Property inference attacks against
  {GANs}. arXiv:2111.07608  (2021)

\end{thebibliography}

\appendix

\section*{Appendix}


\begin{preprint}



\section{Selecting Window Size}\label{app:windowsize}

Ideally, we want to train \attclf to distinguish between \preq and !\preq. We compute the AUC scores under the FAR-TAR/FRR-TRR curve when \attclf is trained to distinguish between \preq and !\preq. We present this in Table~\ref{tab:aucattestation} under window size ``0''. We observe that \attclf has a high AUC for the edge ratios compared to the middle ratios. Hence, \attclf is not effective for all \preq values.

To this end, we relax the requirement to distinguish between \preq and !\preq. Instead, we evaluate if increasing the window size improves the success of \attclf in distinguishing between \{\preq-1,\preq,\preq+1\} and the remaining ratios (!\{\preq-1,\preq,\preq+1\}). This is indicated in Table~\ref{tab:aucattestation} under window size ``$\pm$1''.

\begin{table}[!htb]
\setlength\tabcolsep{2pt}
\scriptsize
\caption{Identifying the best window size (w) using AUC scores to maximize attestation success. \colorbox{red!25}{red} indicates an AUC score $<$0.85 and \colorbox{green!25}{green} indicates an AUC score $>=$0.85. $\pm$1 indicates the window \{$p_{req}$-1,$p_{req}$,$p_{req}$+1\} but only two of those ratios are combined for the edge values. Performance increases on using window size of $\pm$1 instead of 0. \arxiv and \boneage do well across different \preq but \sexcensus and \racecensus have low effectiveness for certain \preq.}
 \begin{subtable}[t]{0.5\linewidth}
      \centering
        \caption{\arxiv}
\begin{tabular}{ c|c|c} 
 \toprule
  \textbf{$p_{req}$}  & \multicolumn{2}{c}{Window Size} \\
 & 0 & $\pm$1\\ 
 \toprule
\textbf{9}   & \cellcolor{green!30}0.99  & \cellcolor{green!30}1.00\\ 
\textbf{10}  & \cellcolor{green!30}0.89 & \cellcolor{green!30}1.00\\ 
\textbf{11}  & \cellcolor{green!30}0.88 & \cellcolor{green!30}0.92\\ 
\textbf{12}  & \cellcolor{red!30}0.77 & \cellcolor{green!30}0.96\\ 
\textbf{13}  & \cellcolor{green!30}0.93 & \cellcolor{green!30}0.93\\ 
\textbf{14}  & \cellcolor{green!30}0.98 & \cellcolor{green!30}0.99\\ 
\textbf{15}  & \cellcolor{green!30}0.99  & \cellcolor{green!30}0.99\\ 
\textbf{16}  & \cellcolor{green!30}1.00 & \cellcolor{green!30}1.00\\ 
\textbf{17}  & \cellcolor{green!30}1.00 & \cellcolor{green!30}1.00\\ 
 \bottomrule
\end{tabular}
    \end{subtable}%
    \begin{subtable}[t]{0.5\linewidth}
      \centering
        \caption{\boneage}
\begin{tabular}{ c|c|c } 
 \toprule
  \textbf{$p_{req}$}  & \multicolumn{2}{c}{Window Size} \\
 & 0 & $\pm$1\\ 
 \toprule
\textbf{0.20} &  \cellcolor{green!30}0.97 & \cellcolor{green!30}0.99\\
\textbf{0.30} & \cellcolor{red!30}0.84 & \cellcolor{green!30}0.99\\ 
\textbf{0.40} & \cellcolor{green!30}0.88 & \cellcolor{green!30}0.92\\
\textbf{0.50} & \cellcolor{red!30}0.80 & \cellcolor{green!30}0.86\\
\textbf{0.60} & \cellcolor{red!30}0.68 & \cellcolor{green!30}0.87\\
\textbf{0.70} &\cellcolor{red!30}0.80 & \cellcolor{green!30}0.99\\
\textbf{0.80} & \cellcolor{green!30}0.97 & \cellcolor{green!30}0.95\\ 
 \bottomrule
\end{tabular}
    \end{subtable}\\
    \begin{subtable}[t]{0.5\linewidth}
      \centering
        \caption{\sexcensus}
\begin{tabular}{ c|c|c } 
 \toprule
  \textbf{$p_{req}$}  & \multicolumn{2}{c}{Window Size} \\
 & 0 & $\pm$1\\ 
 \toprule
\textbf{0.00} & \cellcolor{green!30}1.00 & \cellcolor{green!30}0.93\\ 
\textbf{0.10} & \cellcolor{green!30}0.95 & \cellcolor{red!30}0.81\\
\textbf{0.20} & \cellcolor{red!30}0.74 & \cellcolor{green!30}0.91\\
\textbf{0.30} & \cellcolor{red!30}0.83 & \cellcolor{red!30}0.79 \\
\textbf{0.40} & \cellcolor{red!30}0.64 & \cellcolor{red!30}0.68\\
\textbf{0.50} & \cellcolor{red!30}0.55 & \cellcolor{red!30}0.65\\
\textbf{0.60} & \cellcolor{red!30}0.75 & \cellcolor{red!30}0.63 \\
\textbf{0.70} & \cellcolor{red!30}0.53 & \cellcolor{red!30}0.67\\
\textbf{0.80} & \cellcolor{red!30}0.72 & \cellcolor{green!30}0.88\\ 
\textbf{0.90} & \cellcolor{red!30}0.77 & \cellcolor{green!30}0.89\\
\textbf{1.00} & \cellcolor{green!30}1.00 & \cellcolor{green!30}0.85\\
 \bottomrule
\end{tabular}
    \end{subtable}%
    \begin{subtable}[t]{0.5\linewidth}
      \centering
        \caption{\racecensus}
\begin{tabular}{ c|c|c } 
 \toprule
  \textbf{$p_{req}$}  & \multicolumn{2}{c}{Window Size} \\
 & 0 & $\pm$1\\ 
 \toprule
\textbf{0.00} & \cellcolor{green!30}1.00 & \cellcolor{red!30}0.88\\
\textbf{0.10} & \cellcolor{red!30}0.84 & \cellcolor{green!30}0.89 \\
\textbf{0.20} & \cellcolor{red!30}0.70 & \cellcolor{green!30}0.95\\
\textbf{0.30} & \cellcolor{red!30}0.81 & \cellcolor{red!30}0.71\\
\textbf{0.40} & \cellcolor{red!30}0.67 & \cellcolor{red!30}0.74\\
\textbf{0.50} & \cellcolor{red!30}0.57 & \cellcolor{red!30}0.65\\
\textbf{0.60} & \cellcolor{red!30}0.67 & \cellcolor{red!30}0.63\\
\textbf{0.70} & \cellcolor{red!30}0.76 & \cellcolor{red!30}0.83\\
\textbf{0.80} & \cellcolor{red!30}0.83 & \cellcolor{green!30}0.96\\
\textbf{0.90} & \cellcolor{green!30}0.92 & \cellcolor{red!30}0.83 \\
\textbf{1.00} & \cellcolor{green!30}1.00 & \cellcolor{red!30}0.75\\ 
 \bottomrule
\end{tabular}
    \end{subtable} 
    \label{tab:aucattestation}
\end{table}

We set a threshold of 0.85 for AUC scores where a score $>=0.85$ is considered good and marked in \colorbox{green!25}{green} and \colorbox{red!25}{red} otherwise.
For \arxiv and \boneage, we observe that the window size of $\pm$1 is better than a window size of 0 (visually indicated by more \colorbox{green!25}{green} cells). Hence, for these two datasets, we choose window size of $\pm$1 due to better effectiveness. We attribute this to better distinguishability between models trained on datasets with different property values (as shown in~\cite{suripropinf}).

For \racecensus and \sexcensus, we note that the AUC scores are not high for most ratios in the middle whereas the edge ratios (``0.00'' and ``1.00'') have a perfect score. This suggests that for some \preq values in the middle, the cost for attestation is higher than the edge \preq values. We attribute this to the lower distinguishability for different \preq values (as shown in~\cite{suripropinf}).

\verifier can do a similar analysis beforehand by evaluating on \verifiertestdata to see the potential cost of attestation for different datasets and identify the window sizes for specific \preq values which could improve attestation effectiveness.

\section{Effectiveness of Robust \attclf}\label{app:robustVerdata}

In our earlier discussion, we highlighted the use of AUC scores for FAR-TAR curves to demonstrate that the performance on clean \verificationdata remains comparable whether adversarial training is applied or not to achieve a robust \attclf. To provide a more comprehensive assessment of the effectiveness of our robust \attclf, we present additional metrics in Table~\ref{tab:effectivenessRobust}, including TRR@5\% FRR, TAR@5\% FAR, and EER.

\begin{table}[!htb]
\setlength\tabcolsep{1pt}
\scriptsize
\caption{Robust \attclf effectiveness using TAR, TRR with 5\% thresholds for FAR, FRR along with EER across different \preq windows on \verificationdata. The \preq value within the window is indicated in \textbf{bold}. We observe a similar trend as before: edge \preq values have higher effectiveness than middle \preq values.}
    \begin{subtable}[t]{.5\linewidth}
      \centering
        \caption{\arxiv}
\begin{tabular}{ c|c|c|c} 
 \toprule
 \textbf{\preq Range}  & \textbf{TAR}  &  \textbf{TRR } & \textbf{EER}\\ 
 \toprule
\{\textbf{9}, 10\}  & 0.94 & 0.94 & 0.06 \\ 
\{9, \textbf{10}, 11\} & 0.99 & 0.99 & 0.02 \\ 
\{10, \textbf{11}, 12\} & 0.20 & 0.83 & 0.16\\ 
\{11, \textbf{12}, 13\} & 0.37 & 0.59 & 0.29\\ 
\{12, \textbf{13}, 14\} & 0.72 & 0.83 & 0.11 \\ 
\{13, \textbf{14}, 15\} & 0.86 & 0.85 & 0.10\\ 
\{14, \textbf{15}, 16\} & 0.66 & 0.86 & 0.12\\ 
\{15, \textbf{16}, 17\} & 1.00 & 1.00 & 0.00 \\ 
\{16, \textbf{17}\} & 1.00 & 1.00 & 0.00\\ 
 \bottomrule
\end{tabular}
    \end{subtable}
    \begin{subtable}[t]{.5\linewidth}
      \centering
        \caption{\boneage}
\begin{tabular}{ c|c|c|c } 
 \toprule
 \textbf{\preq Range} & \textbf{TAR}  & \textbf{TRR} & \textbf{EER}\\ 
 \toprule
\{\textbf{0.20}, 0.30\} & 0.97 & 0.97 & 0.05 \\ 
\{0.20, \textbf{0.30}, 0.40\} & 0.99 & 1.00 & 0.02\\ 
\{0.30, \textbf{0.40}, 0.50\} & 0.81 & 0.87 & 0.10\\ 
\{0.40, \textbf{0.50}, 0.60\} & 0.43 & 0.59 & 0.23\\ 
\{0.50, \textbf{0.60}, 0.70\} & 0.26 & 0.72 & 0.25\\ 
\{0.60, \textbf{0.70}, 0.80\} & 0.97 & 0.98 & 0.04\\ 
\{0.70, \textbf{0.80}\} & 0.93 & 0.94 & 0.05\\ 
 \bottomrule
\end{tabular}
    \end{subtable}\\
\begin{subtable}[t]{.5\linewidth}
\centering
 \caption{\sexcensus}
\begin{tabular}{ c|c|c|c} 
 \toprule
 {\textbf{\preq Range}} & 
 {\textbf{TAR}} & 
 {\textbf{TRR}} & 
 {\textbf{EER}}\\ 
 \toprule
 \{\textbf{0.00}\} & 1.00 & 1.00 & 0.00\\ 
\{0.00, \textbf{0.10}, 0.20\} & 0.45 & 0.56 & 0.21\\ 
 \{0.10, \textbf{0.20}, 0.30\} & 0.57 & 0.66 & 0.17\\ 
\{0.20, \textbf{0.30}, 0.40\} & 0.20 & 0.44 & 0.26\\ 
\{0.30, \textbf{0.40}, 0.50\} & 0.11 & 0.29 & 0.38\\ 
\{0.40, \textbf{0.50}, 0.60\} & 0.14 & 0.18 & 0.39\\ 
\{0.50, \textbf{0.60}, 0.70\} & 0.16 & 0.19 & 0.38\\
\{0.60, \textbf{0.70}, 0.80\} & 0.16 & 0.22 & 0.34\\ 
\{0.70, \textbf{0.80}, 0.90\} & 0.55 & 0.58 & 0.19\\ 
\{0.80, \textbf{0.90}, 1.00\} & 0.62 & 0.61 & 0.19\\ 
 \{\textbf{1.00}\} & 1.00 & 1.00 & 0.00\\ 
 \bottomrule
\end{tabular}
    \end{subtable}
    \begin{subtable}[t]{.5\linewidth}
      \centering
        \caption{\racecensus}
\begin{tabular}{ c|c|c|c} 
 \toprule
 {\textbf{\preq Range}} & 
 {\textbf{TAR}} & 
 {\textbf{TRR}} & 
 {\textbf{EER}}\\ 
 \toprule
 \{\textbf{0.00}\} & 1.00 & 1.00 & 0.00\\ 
 \{0.00, \textbf{0.10}, 0.20\} & 0.06 & 0.51 & 0.36 \\
 \{0.10, \textbf{0.20}, 0.30\} & 0.66 & 0.86 & 0.10\\ 
 \{0.20, \textbf{0.30}, 0.40\} & 0.22 & 0.49 & 0.29\\ 
 \{0.30, \textbf{0.40}, 0.50\} & 0.10 & 0.23 & 0.39\\ 
 \{0.40, \textbf{0.50}, 0.60\} & 0.12 & 0.13 & 0.41\\ 
 \{0.50, \textbf{0.60}, 0.70\} & 0.10 & 0.25 & 0.38\\
 \{0.60, \textbf{0.70}, 0.80\} & 0.10 & 0.39 & 0.29\\ 
 \{0.70, \textbf{0.80}, 0.90\} & 0.66 & 0.82 & 0.12\\ 
 \{0.80, \textbf{0.90}, 1.00\} & 0.49 & 0.59 & 0.22\\ 
 \{\textbf{1.00}\} & 1.00 & 1.00 & 0.00\\ 
 \bottomrule
\end{tabular}
    \end{subtable}
\label{tab:effectivenessRobust}
\end{table}

Similar to the outcomes shown in Table 2 for the non-robust \attclf's effectiveness, we observe elevated TRR and TAR values for edge \preq values, but diminished performance for the middle \preq values. This trend can be attributed to differences in distinguishability, a notion underscored by Suri and Evans~\cite{suripropinf}. Additionally, we find that some EER values for specific \preq values outperform both FAR and FRR, suggesting the potential for a superior threshold beyond the standard 5\%.

However, it's important to highlight that despite this improvement, the inference-based attestation utilizing a robust \attclf remains ineffective for certain \preq values. This implies that, despite its robustness, relying solely on inference-based attestation is inadequate.

\section{Impact of Adversarial Finetuning Attack on \modelcheck's Accuracy}\label{app:accfinetuning}

We examined the capabilities of \malprover who employs an adversarial finetuning attack strategy. This involves introducing adversarial noise to the parameters of the first layer, followed by the fine-tuning of subsequent layers. Consequently, \malprover is able to fool \verifier's attestation process by exploiting the adversarially perturbed first layer parameters.


\begin{table}[!htb]
\setlength\tabcolsep{1pt}
\scriptsize
\caption{Average accuracy across 1000 shadow models w/ and w/o malicious \prover modifying the first layer parameters followed by fine-tuning. We refer to them as $\mathcal{A}_{orig}$ and $\mathcal{A}_{att}$ respectively. \colorbox{green!25}{green} indicates $\mathcal{A}_{orig}$ $>>$ $\mathcal{A}_{att}$, \colorbox{orange!25}{orange} indicates $\mathcal{A}_{orig}>\mathcal{A}_{att}$ ($<$2\%) and \colorbox{red!25}{red} indicates $\mathcal{A}_{orig}$ $\sim$ $\mathcal{A}_{att}$. Across most \preq values, malicious \prover can restore $\mathcal{A}_{att}$ close to $\mathcal{A}_{orig}$.}
 \begin{subtable}[t]{0.5\linewidth}
      \centering
        \caption{\arxiv}
\begin{tabular}{ c|c|c} 
 \toprule
  \textbf{$p_{req}$ Range}  & \multicolumn{2}{c}{Shadow Model Accuracy} \\
 & $\mathcal{A}_{orig}$ & $\mathcal{A}_{att}$\\ 
 \toprule
\{\textbf{9}, 10\}  & \cellcolor{orange!30}0.72 $\pm$ 0.00 & \cellcolor{orange!30}0.70 $\pm$ 0.00 \\ 
\{9, \textbf{10}, 11\} & \cellcolor{orange!30}0.72 $\pm$ 0.00 & \cellcolor{orange!30}0.70 $\pm$ 0.00 \\ 
\{10, \textbf{11}, 12\} & \cellcolor{orange!30}0.72 $\pm$ 0.00 & \cellcolor{orange!30}0.71 $\pm$ 0.00\\ 
\{11, \textbf{12}, 13\} & \cellcolor{orange!30}0.72 $\pm$ 0.00 & \cellcolor{orange!30}0.71 $\pm$ 0.00 \\ 
\{12, \textbf{13}, 14\} & \cellcolor{orange!30}0.72 $\pm$ 0.00 & \cellcolor{orange!30}0.71 $\pm$ 0.00\\ 
\{13, \textbf{14}, 15\} & \cellcolor{orange!30}0.72 $\pm$ 0.00 & \cellcolor{orange!30}0.71 $\pm$ 0.00\\ 
\{14, \textbf{15}, 16\} & \cellcolor{red!30}0.72 $\pm$ 0.00 & \cellcolor{red!30}0.72 $\pm$ 0.00\\ 
\{15, \textbf{16}, 17\} & \cellcolor{red!30}0.72 $\pm$ 0.00 & \cellcolor{red!30}0.72 $\pm$ 0.00 \\
\{16, \textbf{17}\} & \cellcolor{red!30}0.72 $\pm$ 0.00 & \cellcolor{red!30}0.72 $\pm$ 0.01 \\ 
 \bottomrule
\end{tabular}
    \end{subtable}%
    \begin{subtable}[t]{0.5\linewidth}
      \centering
        \caption{\boneage}
\begin{tabular}{ c|c|c } 
 \toprule
  \textbf{$p_{req}$ Range}  & \multicolumn{2}{c}{Shadow Model Accuracy} \\
 & $\mathcal{A}_{orig}$ & $\mathcal{A}_{att}$\\ 
 \toprule
\{\textbf{0.20}, 0.30\} & \cellcolor{red!30}0.79 $\pm$ 0.02 & \cellcolor{red!30}0.79 $\pm$ 0.02\\ 
\{0.20, \textbf{0.30}, 0.40\} & \cellcolor{red!30}0.78 $\pm$ 0.02 & \cellcolor{red!30}0.76 $\pm$ 0.02\\ 
\{0.30, \textbf{0.40}, 0.50\} & \cellcolor{red!30}0.78 $\pm$ 0.02 & \cellcolor{red!30}0.75 $\pm$ 0.02\\ 
\{0.40, \textbf{0.50}, 0.60\} & \cellcolor{green!30}0.78 $\pm$ 0.02 & \cellcolor{green!30}0.70 $\pm$ 0.02\\ 
\{0.50, \textbf{0.60}, 0.70\} & \cellcolor{red!30}0.78 $\pm$ 0.02 & \cellcolor{red!30}0.78 $\pm$ 0.01\\ 
\{0.60, \textbf{0.70}, 0.80\} & \cellcolor{red!30}0.78 $\pm$ 0.02 & \cellcolor{red!30}0.78 $\pm$ 0.01\\ 
\{0.70, \textbf{0.80}\} &  \cellcolor{red!30}0.78 $\pm$ 0.01 & \cellcolor{red!30}0.79 $\pm$ 0.01 \\ 
 \bottomrule
\end{tabular}
    \end{subtable}\\
    \begin{subtable}[t]{0.5\linewidth}
      \centering
        \caption{\sexcensus}
\begin{tabular}{ c|c|c } 
 \toprule
  \textbf{$p_{req}$ Range}  & \multicolumn{2}{c}{Shadow Model Accuracy} \\
 & $\mathcal{A}_{orig}$ & $\mathcal{A}_{att}$\\ 
 \toprule
 \{\textbf{0.00}\} & \cellcolor{red!30}0.77 $\pm$ 0.04 & \cellcolor{red!30}0.80 $\pm$ 0.01\\ 
 \{0.00, \textbf{0.10}, 0.20\} & \cellcolor{red!30}0.79 $\pm$ 0.03 & \cellcolor{red!30}0.82 $\pm$ 0.01\\
 \{0.10, \textbf{0.20}, 0.30\} & \cellcolor{red!30}0.79 $\pm$ 0.03 & \cellcolor{red!30}0.82 $\pm$ 0.01\\ 
 \{0.20, \textbf{0.30}, 0.40\} & \cellcolor{red!30}0.80 $\pm$ 0.02 & \cellcolor{red!30}0.82 $\pm$ 0.01\\ 
 \{0.30, \textbf{0.40}, 0.50\} & \cellcolor{red!30}0.81 $\pm$ 0.02 & \cellcolor{red!30}0.82 $\pm$ 0.01\\ 
 \{0.40, \textbf{0.50}, 0.60\} & \cellcolor{red!30}0.82 $\pm$ 0.02 & \cellcolor{red!30}0.83 $\pm$ 0.01\\ 
 \{0.50, \textbf{0.60}, 0.70\} & \cellcolor{red!30}0.82 $\pm$ 0.02 & \cellcolor{red!30}0.84 $\pm$ 0.01 \\
 \{0.60, \textbf{0.70}, 0.80\} & \cellcolor{red!30}0.82 $\pm$ 0.04 & \cellcolor{red!30}0.84 $\pm$ 0.01\\ 
 \{0.70, \textbf{0.80}, 0.90\} & \cellcolor{red!30}0.83 $\pm$ 0.02 & \cellcolor{red!30}0.84 $\pm$ 0.01 \\ 
 \{0.80, \textbf{0.90}, 1.00\} & \cellcolor{red!30}0.85 $\pm$ 0.05 & \cellcolor{red!30}0.86 $\pm$ 0.01 \\ 
 \{\textbf{1.00}\} & \cellcolor{red!30}0.87 $\pm$ 0.03 & \cellcolor{red!30}0.88 $\pm$ 0.01\\ 
 \bottomrule
\end{tabular}
    \end{subtable}%
    \begin{subtable}[t]{0.5\linewidth}
      \centering
        \caption{\racecensus}
\begin{tabular}{ c|c|c } 
 \toprule
  \textbf{$p_{req}$ Range}  &  \multicolumn{2}{c}{Shadow Model Accuracy} \\
 & $\mathcal{A}_{orig}$ & $\mathcal{A}_{att}$\\ 
 \toprule
 \{\textbf{0.00}\} & \cellcolor{red!30}0.87 $\pm$ 0.04 & \cellcolor{red!30}0.87 $\pm$ 0.01 \\ 
 \{0.00, \textbf{0.10}, 0.20\} & \cellcolor{red!30}0.86 $\pm$ 0.03 & \cellcolor{red!30}0.87 $\pm$ 0.01 \\
 \{0.10, \textbf{0.20}, 0.30\} & \cellcolor{red!30}0.86 $\pm$ 0.02 & \cellcolor{red!30}0.86 $\pm$ 0.01\\ 
 \{0.20, \textbf{0.30}, 0.40\} & \cellcolor{red!30}0.85 $\pm$ 0.01 & \cellcolor{red!30}0.85 $\pm$ 0.01 \\ 
 \{0.30, \textbf{0.40}, 0.50\} & \cellcolor{red!30}0.84 $\pm$ 0.02 & \cellcolor{red!30}0.85 $\pm$ 0.01 \\ 
 \{0.40, \textbf{0.50}, 0.60\} & \cellcolor{red!30}0.84 $\pm$ 0.02 & \cellcolor{red!30}0.85 $\pm$ 0.01 \\ 
 \{0.50, \textbf{0.60}, 0.70\} & \cellcolor{red!30}0.83 $\pm$ 0.02 & \cellcolor{red!30}0.84 $\pm$ 0.01 \\
 \{0.60, \textbf{0.70}, 0.80\} & \cellcolor{red!30}0.82 $\pm$ 0.02 & \cellcolor{red!30}0.83 $\pm$ 0.01\\ 
 \{0.70, \textbf{0.80}, 0.90\} & \cellcolor{red!30}0.81 $\pm$ 0.02 & \cellcolor{red!30}0.82 $\pm$ 0.01 \\ 
  \{0.80, \textbf{0.90}, 1.00\} & \cellcolor{red!30}0.81 $\pm$ 0.02 & \cellcolor{red!30}0.82 $\pm$ 0.01 \\ 
 \{\textbf{1.00}\} & \cellcolor{red!30}0.81 $\pm$ 0.02 & \cellcolor{red!30}0.82 $\pm$ 0.01 \\ 
 \bottomrule
\end{tabular}
    \end{subtable} 
    \label{tab:accuracy}
\end{table}

We indicate the impact of perturbing the first layer parameters on the \malprover's \modelcheck accuracy in Table~\ref{tab:accuracy}. We observe that adding noise to first layer followed by fine-tuning restores the accuracy of model ($\mathcal{A}_{att}$) close to the original accuracy ($\mathcal{A}_{orig}$) while still being able to fool the attestation. Hence, the attack can be done without a significant decrease in accuracy.
\end{preprint}

\section{Details for Cryptographic Attestation}\label{app:crypto}

\Hypar{Protocol Instantiation} Given the proof objectives, our goal now is to ﬁnd a concrete cryptographic protocol variant to instantiate property attestation. To this end, we need the following: (1) primitives for ML training, (2) security against \malprover, and (3) an efﬁcient protocol instantiation that allows us to use the cryptographic property attestation in a real setting. 
We rule out TEEs because of their susceptibility to side-channel attacks, hence violating (2). Because of their impracticality to deploy in real-world sized models, thus violating (3), we also rule out Homomorphic Encryption (HE)~\cite{sokSecureInf}. As of now, there are efficient ZKPs for verifiable inference~\cite{mystique}, but not for backpropagation during ML training, violating (1) and ruling out ZKPs for our instantiation. As state-of-the-art works in PPML based on MPC such as CrypTen~\cite{knott2021crypten} satisfy all three required properties for our cryptographic property attestation, we choose MPC as instantiation.

\Hypar{Property Attestation as a Cryptographic Protocol} 
Assuming a malicious \prover and semi-honest \verifier, we construct a cryptographic protocol based on MPC in the outsourcing setting with two non-colluding semi-honest servers \serverone and \servertwo. 
The protocol consists of the following steps:
\begin{enumerate}[leftmargin=*]
    \item Initiate input-sharing phase between \prover and \serverone, \servertwo.
    \item \serverone and \servertwo run \dc on their input shares of \proverdata.
    \item \serverone and \servertwo securely train on their input shares, which yields \modelmpc
    \item \serverone and \servertwo send output shares of \dc and \modelmpc to \verifier for reconstruction of plaintext outputs. 
    \item \verifier checks if \dc succeeded using the output shares.
\end{enumerate}

\Itpar{(1) Input-sharing Phase} \prover computes additive secret-shares of the training dataset (\proverdata).  Hence, the prover computes $\proverdatashare{1}$, $\proverdatashare{2}$ such that $\proverdatashare{1} + \proverdatashare{2} = \mathcal{D}_{\mathcal{P}}^{tr}$ and sends $\proverdatashare{1}$ to \serverone and $\proverdatashare{2}$ to \servertwo. 

\Itpar{(2) Secure Computation of \dc} Given the input shares of \proverdata, \serverone and \servertwo compute \dc by computing the distributional property of \proverdata and comparing against \preq.

\Itpar{(3) Secure Training of \modelmpc} Given the input shares of both \proverdata and \modelcheck, both servers jointly run the protocols for secure training as described in~\cite{knott2021crypten}. CrypTen has efficient secure protocols for both the forward pass and back propagation. We refer to~\cite{knott2021crypten} for the protocol details. We emphasize that \serverone and \servertwo use the previously obtained shares of \proverdata from the input-sharing phase, because they were used for \dc. This leaves no room for \prover to cheat by choosing different shares of another dataset $\mathcal{D}' \neq \mathcal{D}^{tr}_{\mathcal{P}}$ for training. 

\Itpar{(4) Verify \dc} \serverone and \servertwo send the output shares $[v]^1$ and $[v]^2$ of \dc to \verifier who now locally reconstructs the output $v = [v]^1 + [v]^2$. Now, $v=1$ iff \dc was successful. Then, \verifier reconstructs \modelmpc by locally adding the output shares from \serverone and \servertwo, i.e., $\mathcal{M}_{2pc} = \modelmpcshare{1} + \modelmpcshare{2}$. 

\makeparafit
\Hypar{Security and Correctness of Cryptographic Attestation}
Since we implement cryptographic attestation using CrypTen, we refer to~\cite{knott2021crypten} for the detailed security proofs. Assuming CrypTen's protocols satisfy security and correctness, we discuss the security (i.e., preserving the privacy of \prover's dataset \proverdata) and correctness for cryptographic attestation.

\makeparafit
The privacy of \proverdata naturally follows from the security guarantees of linear secret-sharing~\cite{GMW87,knott2021crypten}. 
For correctness, we identify two cases: 
\begin{itemize}[leftmargin=*]
\item \textbf{when \prover does not cheat} and correctly creates an input sharing for \proverdata, then, correctness follows from the underlying secret-sharing scheme.
\item \textbf{when \prover cheats}, then \prover can
\begin{itemize}[leftmargin=*]
\item simply abort instead of providing a valid sharing to escape attestation, hence the whole attestation fails.
\item create incorrect shares $\maliciousdatashare{1}$ and $\maliciousdatashare{2}$ where $\maliciousdatashare{1} + \maliciousdatashare{2} = \mathcal{D}' \neq \mathcal{D}_\mathcal{P}$. However, since the two shares indeed compute $\mathcal{D}'$, the input-sharing is done correctly, just for a different input value. MPC does not secure against choosing the ``wrong'' input value. However, this is not a problem, because if $\mathcal{D}'$ satisfies \preq, we still obtain a valid model with respect to the distributional property. 
\end{itemize}
\end{itemize} 
After the input-sharing phase, \prover does not participate in the protocol. Hence, there is no further cheating as \serverone, \servertwo, \verifier are semi-honest. Secure computation of~\dc only consists of secure additions and comparison, hence the correctness and privacy of~\dc as well as secure training of~\modelmpc directly follows from the security guarantees of CrypTen~\cite{knott2021crypten}.




\begin{preprint}
\section{Hybrid Attestation: FAR vs. Cost on \verificationdata}
\label{sec:hybridverdata}

For hybrid attestation, we showed the trade-off between FAR and expected cost on varying \Nspotcheck for \verifiertestdata. \verifier uses this to identify the suitable value of \Nspotcheck.

To get the actual estimate of FAR and expected cost, \verifier can use the chosen value of \Nspotcheck in a plot between FAR and expected cost in Figure~\ref{fig:farVcostVerify}.

\begin{figure}[!htb]
\begin{minipage}[h]{0.49\linewidth}
\begin{center}
\includegraphics[width=0.8\linewidth]{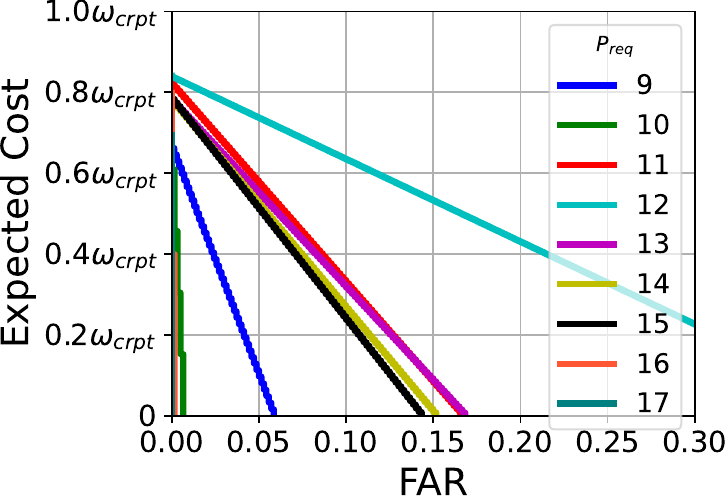} 
\subcaption{\arxiv}
\end{center} 
\end{minipage}
\hfill
\begin{minipage}[h]{0.49\linewidth}
\begin{center}
\includegraphics[width=0.8\linewidth]{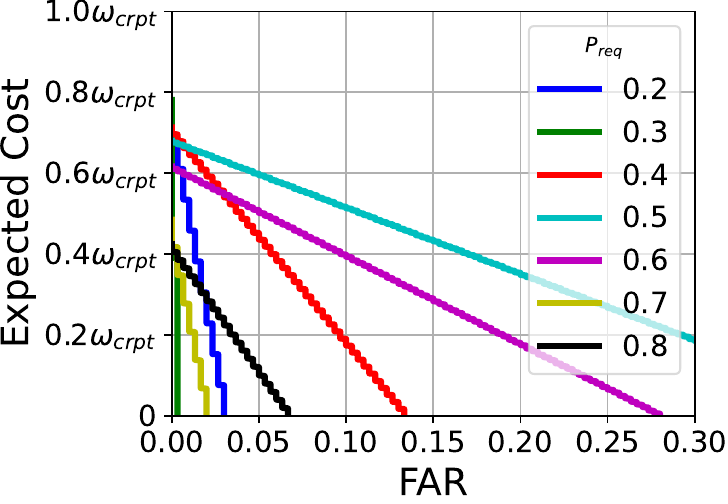} 
\subcaption{\boneage}
\end{center}
\end{minipage}
\vfill
\begin{minipage}[h]{0.49\linewidth}
\begin{center}
\includegraphics[width=0.8\linewidth]{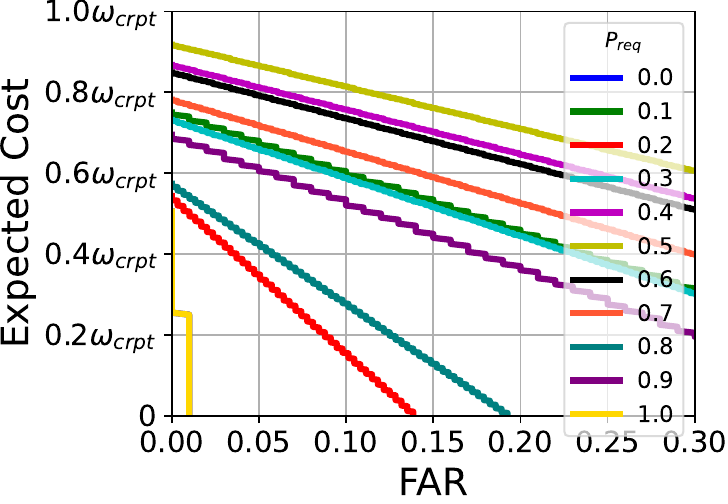} 
\subcaption{\racecensus}
\end{center}
\end{minipage}
\hfill
\begin{minipage}[h]{0.49\linewidth}
\begin{center}
\includegraphics[width=0.8\linewidth]{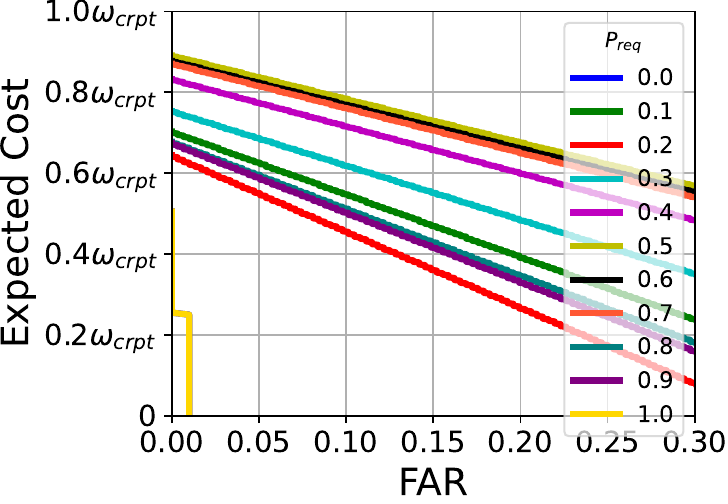} 
\subcaption{\sexcensus}
\end{center}
\end{minipage}
\caption{Fixed FRR@5\%: Expected cost on varying \Nspotcheck on \verificationdata. \verifier can use suitable value of \Nspotcheck identified from \verifiertestdata to note the corresponding FAR and expected cost.}
\label{fig:farVcostVerify}
\end{figure}

We still observe that expected cost is less for hybrid attestation than for cryptographic attestation for all accepted \prover{s} and FAR is less than no spot-checks (FAR same as inference-based attestation).
\end{preprint}

\end{document}